\documentclass[a4paper,8pt,onecolumn]{quantumarticle}
\pdfoutput=1
\usepackage{amsmath, amssymb, amsthm}
\usepackage{graphics}
\usepackage{hyperref}
\usepackage{orcidlink}
\usepackage{bm,subcaption}

\newtheorem{theorem}{Theorem}
\newtheorem{lemma}{Lemma}[section]
\newtheorem*{claim}{Statement}
\newtheorem*{remark}{Remark}
\newtheorem{corollary}{Corollary}

\begin{document}
\title[Structure theorem for quantum limit cycle attractors]{Structure theorem for quantum limit cycle attractors: A geometric path-integral approach}
\author{Nana Siddhartha\,\orcidlink{0000-0003-4257-4611}}
\date{24 November 2026}

\affiliation{Institute of Physics, Slovak Academy of Sciences, Dúbravská cesta 9, 845 11 Bratislava, Slovakia}
\affiliation{Department of Theoretical Physics, Comenius University, Bratislava, Slovakia}
\affiliation{Chennai Mathematical Institute, SIPCOT IT Park, Siruseri 603103, India}

\email{nana.siddhartha@savba.sk}

\begin{abstract}
We show a Structure Theorem for an arbitrary separable quantum system (\(\dim\mathcal{H}\leq\aleph_0\)) that is coupled, through a symmetric bidirectional phase interface, to a self-sustaining phase-oscillator environment. Under the  assumptions of \(U(1)\) covariance, instantaneous Lindblad bounded positivity, and the existence of an entrainable periodic attractor, the reduced generator is forced into a universal operator structure consisting of a state-dependent coherent recoil together with single-quantum feedback. Positivity further imposes a sharp dimensional dichotomy: purely dissipative gain saturation is forbidden in finite dimensions by trace-norm contraction, so coherent feedback is mandatory for qubits and qudits, whereas unbounded ladder algebras permit purely dissipative saturation.
Expanding the general feedback functionals to leading order in the coupling recovers a canonical polynomial master equation that remains completely positive and trace-preserving in every dimension and combines additively with ordinary linear Lindblad channels. An accompanying Inverse Coupling Theorem shows that an autonomous periodic orbit can be sustained only when the interaction manifold has non-trivial fundamental group \(\pi_1(\mathcal{M})=\mathbb{Z}\), thereby isolating the sinusoidal \(U(1)\)-covariant phase coupling as the generic mechanism. The resulting implementation is realized explicitly for a single qubit, a bipartite two-qubit system, and a bosonic Fock space.
\end{abstract}
\maketitle

\section{Introduction}
\label{sec:intro}

Self-sustained oscillations, or limit cycles, are ubiquitous across nonlinear dynamical systems, governing everything from biochemical clocks and neural firing to laser cavities and coupled mechanical oscillators \cite{Strogatz2015,Pikovsky2001}. Unlike thermal equilibria, a limit cycle is an autonomous periodic attractor---a closed trajectory in phase space that absorbs perturbations and maintains a fixed amplitude and frequency without external periodic driving. Extending this foundational phenomenon into the quantum regime, however, exposes a deep structural tension between the unitary, phase-space-preserving dynamical generators of quantum mechanics and the non-conservative phase-space contraction required to stabilize a periodic orbit \cite{Lee2013,Walter2014}.

Open quantum systems governed by time-independent, strictly contractive GKSL generators are known to relax to stationary states. Linear multi-quantum models can at best produce phase-diffused rings of vanishing coherence; they cannot sustain a phase-coherent limit cycle that autonomously entrains to, or phase-locks with, an external oscillator. Systems capable of genuine self-oscillation and mutual entrainment with a $U(1)$ phase oscillator have therefore lacked a structural classification analogous to the GKSL form.

In standard open quantum systems theory, dissipation and decoherence are modeled either via the Gorini-Kossakowski-Sudarshan-Lindblad (GKSL) master equation \cite{Gorini1976,Lindblad1976} or through path-integral treatments such as the Caldeira--Leggett influence functional \cite{Caldeira1983,Feynman1963}. Existing models of quantum synchronization typically engineer limit cycles by inserting competing linear single-photon gain $\mathcal{D}_{a^\dagger}[\rho]$ alongside nonlinear two-photon loss $\mathcal{D}_{a^2}[\rho]$ \cite{Lee2013,Chia2020}. Because a time-independent linear GKSL semigroup is strictly contractive in the trace norm, such models relax to a \emph{stationary, phase-diffused} steady state ($\dot{\rho}_{\mathrm{ss}} = 0$, $\langle a \rangle = 0$) rather than a phase-coherent, time-periodic density-matrix orbit ($\langle L_-(t) \rangle = R_* e^{-i\Omega_0 t}$). Moreover, relying on composite multi-photon scattering channels leaves open a fundamental question: \textit{what canonical algebraic form must govern the reduced dynamics of a phase-coherent, entrainable quantum limit cycle across both finite and infinite dimensions, and what structural constraints does this attractor impose on the system-environment coupling?}

We utilize the geometric path-integral framework, based on a Poincar\'e-Cartan-like 2-form, introduced in \cite{Siddhartha2026} to formulate the reduced dynamics without assuming a state-dependent template from the outset. Evaluating this action over an $\epsilon$-filament hypersurface for an arbitrary quantum system coupled to a self-oscillating phase environment collapses the continuous bath propagator into discrete integer winding jumps. This topological sector collapse yields a universal, state-dependent master equation governed by $O(\lambda^2)$ coherent bath recoil and single-quantum gain saturation rather than artificial multi-photon decay.

\paragraph*{Scope}
To be clear, we do not seek to explain the microscopic origin of self-sustained oscillations in the environment, nor do we claim to derive how a limit cycle emerges spontaneously in the background environment. Instead, we take the existence of a self-oscillating phase environment as given and ask a structural question: \textit{What canonical algebraic form must the reduced master equation and interaction coupling take when a quantum system entrains to such an oscillator?} While our Inverse Coupling Theorem demonstrates that sustaining an autonomous periodic attractor strictly restricts the topology of the interaction manifold---the environmental limit set---to $\pi_1(\mathcal{M}) = \mathbb{Z}$, investigating the fundamental origin of these background limit cycles lies beyond the scope of the present work.

\paragraph*{\textbf{Outline}}
The remainder of this article is organized as follows. In the Claim below, we summarize our central structural and topological results. A survey of the current results in existing literature and their relevance to our investigation is noted in Section~\ref{sec:lit_review}.   In Section~\ref{sec:setup}, we derive the reduced master equation from the 2-form Keldysh path integral for an arbitrary-dimensional bipartite system coupled to a phase environment, including both the $O(\lambda^2)$ coherent recoil and the trace-preserving Normalization Shield, and identify its positive-rate invariance basin. Section~\ref{sec:inverse_topology} proves the Inverse Coupling Theorem, showing via angular-momentum selection rules why higher coupling harmonics decouple and why simply connected interaction manifolds force trajectories onto stationary foci. In Section~\ref{sec:structure}, we formalize the axiomatic class of entrainable quantum phase oscillators, prove the unified Structure Theorem and its finite- vs.\ infinite-dimensional dichotomy, and extract the leading-order Keldysh truncation and higher-cumulant hierarchy as a corollary. Section~\ref{sec:defense} addresses positivity and causality by establishing frozen-state Lindblad positivity, resolving Pechukas's initial-correlation constraint, and verifying mean-field no-signaling. In Section~\ref{sec:correspondence}, we present explicit finite-dimensional realizations (single-qubit limit cycles, bipartite Kuramoto phase locking, and Arnold-tongue entrainment) alongside the bosonic Fock-space correspondence with phenomenological synchronization models. Finally, Section~\ref{sec:conclusions} summarizes our results and outlines future directions, while Appendix~\ref{app:domain_regularity} provides the functional-analytic lemmas guaranteeing domain regularity, Nagumo invariance, and Lie--Trotter channel additivity for unbounded operators.

\section*{Claim}
\label{sec:claim}

Linear open quantum systems are characterized fully by complete positivity, which restricts every Markovian master equation to the standard GKSL form; nonlinear quantum dynamics has lacked a comparable structural principle. Because general nonlinear operators on Hilbert spaces can take endlessly varied mathematical forms, identifying which equations genuinely represent physical, entrainable quantum phase oscillators requires clear symmetry, bidirectional interface, and attractor stability criteria. We state our central claim:

\begin{claim}[Structure of  Quantum Phase Oscillators]
Let $\mathcal{H}$ be a separable Hilbert space of finite or countably infinite dimension ($\dim\mathcal{H} \leq \aleph_0$), and let $\mathcal{L}$ be a continuous, trace-preserving dynamical generator satisfying joint $U(1)$ equivariance, a symmetric bidirectional phase interface within the frozen-state CPTP Lindblad cone, and an entrainable hyperbolic $S^1$ limit cycle that persists in the uncoupled limit $Z_0 \to 0$ (Axioms~(P1)--(P3)). Then $\mathcal{L}$ generates a globally well-defined, trace-preserving, positivity-preserving limit cycle if and only if:
\begin{enumerate}
\item[(I)] \textbf{Universal Operator Form and Dimensional Dichotomy:} Up to unitary equivalence and phase-covariant linear Lindblad channels $\mathcal{L}_{\mathrm{lin}}$, the generator takes the exact form summed over integer phase-charge sectors $q \in \mathbb{N}$,
\begin{equation}
\mathcal{L}(\rho) = -i\Big[H_0 + \sum_{q \geq 1} h_q(|z_q|^2)Q +\Delta(\rho) + H_{\mathrm{ent}}(t),\, \rho\Big] + \sum_{q \geq 1}\gamma_q^+(|z_q|^2)\mathcal{D}_{L_+^{(q)}}[\rho] + \sum_{q \geq 1}\gamma_q^-(|z_q|^2)\mathcal{D}_{L_-^{(q)}}[\rho] + \mathcal{L}_{\mathrm{lin}}[\rho],
\label{eq:canonical}
\end{equation}
where $z_q(\rho) = \mathrm{Tr}(\rho L_-^{(q)})$, $\Delta(\rho) = \sum_{q \geq 1}\big(g_q(|z_q|^2)z_q L_+^{(q)} + \mathrm{h.c.}\big)$ is the state-dependent coherent recoil, and $H_{\mathrm{ent}}(t) = \hbar\lambda\sum_{q \geq 1}\big(Z_0(t)^q L_-^{(q)} + \mathrm{h.c.}\big)$ is the phase-entrainment bridge. Density-matrix positivity enforces a sharp dimensional split:
\begin{itemize}
\item For $\dim\mathcal{H} < \infty$, purely dissipative gain saturation ($\Delta(\rho) = 0$) is strictly forbidden; sustaining a limit cycle requires active coherent recoil ($\mathrm{Im}\,g_q(0) < 0$) paired with population-inverting pumping ($\gamma_q^+(0) > \gamma_q^-(0)$).
\item For $\dim\mathcal{H} = \aleph_0$ with $[L_-^{(q)}, L_+^{(q)}] > 0$, purely dissipative saturation is permitted whenever $\Delta\gamma_q(s) := \gamma_q^+(s) - \gamma_q^-(s)$ satisfies $\Delta\gamma_q(0) > 0$, $\Delta\gamma_q(R_*^2) = 0$, and $\Delta\gamma_q'(R_*^2) < 0$.
\end{itemize}
Expanding the feedback functionals to leading order in the Keldysh coupling $\lambda$ yields the canonical polynomial master equation ($g_q = \lambda^2 \chi_{0,q}^*$, $\gamma_q^+ = \Phi_0^{(q)} - \Phi_1^{(q)}|z_q|^2$, $\gamma_q^- = \Gamma_2^{(q)}$, $h_q = 0$).

\item[(II)] \textbf{Topological Necessity:} In both finite and infinite dimensions, destabilizing the trivial origin requires $\gamma_q^+(0) > \gamma_q^-(0) > 0$. Consequently, any geometric two-form path-integral realization of $\mathcal{L}$ acts over an environmental interaction manifold $\mathcal{M}$ possessing a non-trivial first homotopy group $\pi_1(\mathcal{M}) = \mathbb{Z}$, uniquely selecting the compact circle group $S^1 \cong U(1)$ and excluding simply connected manifolds.
\end{enumerate}
The complete formulation, dimensional no-go dichotomy, leading-order Keldysh corollary, and infinite-dimensional domain verification are established in Theorem~\ref{thm:structure} of Section~\ref{sec:structure} and Appendix~\ref{app:domain_regularity}.
\end{claim}
\section{Literature Survey and Context}
\label{sec:lit_review}

The study of quantum limit cycles and quantum synchronization has matured rapidly over the past decade, evolving along several complementary theoretical and experimental lines. The earliest and most widely adopted paradigm models self-sustained oscillators within bosonic Fock space using time-independent, linear GKSL master equations---most notably the quantum van der Pol oscillator, which pairs linear single-photon gain $\mathcal{D}_{a^\dagger}[\rho]$ with nonlinear two-photon loss $\mathcal{D}_{a^2}[\rho]$~\cite{Lee2013,Walter2014}. Recent work has substantially broadened this top-down approach, extending it to topological lattices of coupled van der Pol oscillators~\cite{Waechtler2023}, twin limit-cycle Li\'enard systems supporting coexisting attractors~\cite{Kehrer2025}, and general planar nonlinear flows constructed via quadrature-dependent jump operators and cascade quantization~\cite{Chia2020,Chia2025}. Experimentally, mutual synchronization between two engineered quantum van der Pol oscillators has recently been observed in trapped-ion motional modes~\cite{Liu2026}. Yet because these generators are strictly linear and time-independent, trace-norm contractivity pulls the ensemble density matrix toward a stationary, phase-diffused steady state ($\dot{\rho}_{\mathrm{ss}} = 0$, $\langle a \rangle = 0$) whose Wigner function forms a static ring. Dynamical oscillations do not persist at the level of $\rho(t)$ itself, appearing instead in joint correlation functions~\cite{Liu2026} or conditionally along individual stochastic trajectories under continuous heterodyne monitoring~\cite{Nadolny2026}. Moreover, stabilizing the oscillation amplitude within a linear GKSL template relies on engineering multi-photon jump channels ($\mathcal{D}_{a^2}$ or polynomial operators $L_k(\hat{x},\hat{p})$) rather than deriving the canonical single-quantum structure from a microscopic phase-oscillator environment.

A parallel body of work has examined the minimal Hilbert-space dimension capable of supporting a quantum limit cycle. Within time-independent linear GKSL dynamics, Roulet and Bruder~\cite{Roulet2018} showed that a single two-level system ($d = 2$) cannot host a valid limit cycle, identifying a spin-1 ($d = 3$) system as the smallest synchronizable quantum unit. Subsequent theoretical~\cite{ParraLopez2020} and trapped-ion experimental~\cite{Zhang2023} investigations demonstrated phase-locking signatures in a dissipated qubit under an external microwave drive, while related studies have mapped out synchronization blockades and dissipative phase locking in locally interacting spin lattices~\cite{Cabot2023}. However, within linear CPTP semigroups, finite-dimensional trace-norm contraction strictly precludes an autonomous, phase-coherent periodic orbit in $d = 2$ once the external drive is removed ($Z_0 \to 0$), blurring the boundary between genuine autonomous entrainment and passive forced oscillation. As established in Theorem~\ref{thm:structure}, this obstruction is not specific to qubits: purely dissipative gain saturation is forbidden in \emph{any} finite dimension ($\dim\mathcal{H} < \infty$), whereas including the $O(\lambda^2)$ coherent bath recoil $\Delta(\rho)$ naturally sustains globally positive, entrainable limit cycles even in a single qubit.

Persistent, phase-coherent oscillations of the density matrix itself have been explored extensively in the context of boundary and continuous time crystals~\cite{Iemini2018,Zaletel2023} as well as dynamical symmetries of open many-body systems~\cite{Buca2019,Buca2022}. Recent advances in this direction include measurement-induced continuous time crystals~\cite{Krishna2023}, nonequilibrium transitions between distinct dissipative time-crystalline phases~\cite{Cosme2023,Cabot2024b}, chimera and cluster synchronization outside the symmetric subspace~\cite{Solanki2024}, and analyses of their thermodynamic and metrological performance~\cite{Montenegro2023,Cabot2024a,Carollo2024}. While these frameworks demonstrate genuine non-stationary quantum evolution, they operate in a different structural regime from a single-system limit-cycle attractor: boundary and continuous time crystals rely on the thermodynamic many-body limit ($N \to \infty$) to close the Liouvillian gap---decaying at any finite $N$---whereas linear dynamical symmetries~\cite{Buca2019,Buca2022} inhabit fine-tuned decoherence-free subspaces whose oscillation amplitude depends on initial conditions rather than converging to an isolated, hyperbolic attractor ($\mathrm{Re}\,\lambda_{\mathrm{rad}} < 0$) that absorbs radial perturbations.

Complementing these dynamical models, recent diagnostic methods such as Lie-algebraic quantum phase reduction~\cite{Setoyama2024}---including its extension to unbiased heterodyne detection~\cite{Setoyama2025}---and measurement-based trajectory frameworks~\cite{Nadolny2026} provide powerful tools for extracting effective phase-response curves and synchronization measures from open quantum evolutions. Because these techniques are perturbative and diagnostic, they take a phenomenological master equation or stochastic unraveling as input rather than classifying the admissible algebraic form of the reduced generator or constraining the topology of the synchronizing environment.

Taken together, these developments highlight a clear structural gap in the theory of open quantum oscillators: linear GKSL models produce stationary phase-diffused rings or non-hyperbolic subspaces, finite-dimensional spins encounter a linear-CPTP contraction barrier, many-body time crystals require $N \to \infty$, and phase-reduction methods presuppose the underlying master equation. By deriving the reduced dynamics from a 2-form Keldysh path integral over a $U(1)$ phase interface, the present work addresses this gap directly through two complementary results: the Inverse Coupling Theorem (Theorem~\ref{thm:inverse_topology}), which proves that $\pi_1(\mathcal{M}) = \mathbb{Z}$ is topologically necessary to sustain an autonomous limit cycle, and the Unified Structure Theorem (Theorem~\ref{thm:structure}), which classifies all entrainable phase-oscillator generators across $\dim\mathcal{H} \leq \aleph_0$, resolves the finite- versus infinite-dimensional dichotomy, and recovers phenomenological van der Pol and semiclassical laser models as explicit limiting cases.

\section{Setup}
\label{sec:setup}
\subsection{Geometric Setup and the Fundamental Limit-Cycle Master Equation}

\subsubsection{Environmental Topology and Winding Invariants}

A quantum theory of periodic attractors needs to treat the kinematic structure of the surrounding environment correctly. When an open quantum system settles into a stable, self-sustained limit cycle far from equilibrium, it cannot be interacting with a translation-invariant thermal reservoir. A persistent periodic attractor implies that the macroscopic degrees of freedom of the non-equilibrium environment $\rho_E(t)$ have collapsed onto a one-dimensional periodic attractor---topologically equivalent to a circle $S^1 \cong U(1)$ parameterized by a compact phase coordinate $x \in [0, 2\pi)$. We consider an arbitrary-dimensional density matrix $\rho(t) \in \mathcal{S}(\mathcal{H}_{\mathrm{bi}})$ representing the state of a composite bipartite quantum system. The environment state $\rho_E(t)$ is supported natively on this $U(1)$-covariant limit cycle and is spanned by its discrete angular-momentum (Fourier) eigenbasis $\{|n\rangle\}_{n \in \mathbb{Z}}$, where the phase-shift operators act as ladder generators $e^{\pm ix}|n\rangle = |n \pm 1\rangle$ and free environmental evolution acts diagonally via $U_E(\tau)|n\rangle = \alpha_n(\tau)|n\rangle$.

The phase coherence of the unperturbed environmental limit cycle is characterized by its complex order parameter,
\begin{equation}
Z_0(t) = \mathrm{Tr}_E\big(\rho_E(t) e^{ix}\big) = R_E(t) e^{i\Psi_E(t)},
\end{equation}
where $R_E(t) \ge 0$ is the environmental coherence amplitude and $\Psi_E(t) \in [0, 2\pi)$ tracks the global background phase.

The system couples to this phase channel through a compact $U(1)$ sinusoidal phase bridge with coupling strength $\lambda$. From the phase-winding correlation functions of the environmental oscillator over its decay envelope, we define the three non-equilibrium environmental functionals governing the reduced dynamics:
\begin{align}
\Phi_0[\rho_E(t)] &= \lambda^2 \int_0^\infty d\tau\,\mathrm{Tr}_E\!\left[\rho_E(t)\,e^{ix(\tau)}e^{-ix(0)}\right], \\
\chi_0[\rho_E(t)] &= -i \int_0^\infty d\tau\,\mathrm{Tr}_E\!\left[\rho_E(t)\big[e^{ix(\tau)}, e^{-ix(0)}\big]\right]_{\mathrm{conn}}, \\
\Phi_1[\rho_E(t)] &= 2\lambda^4 \int_0^\infty d\tau_1 \int_0^{\tau_1} d\tau_2 \int_0^{\tau_2} d\tau_3\,\mathrm{Tr}_E\!\left[\rho_E(t)\mathcal{T}_{\mathcal{C}}\!\left(e^{ix(\tau_1)}e^{-ix(\tau_2)}e^{ix(\tau_3)}e^{-ix(0)}\right)\right]_{\mathrm{conn}},
\end{align}
where $\mathcal{T}_{\mathcal{C}}$ is the contour time-ordering operator and the subscript ``conn'' denotes the fully connected part of the winding correlation function. Here, $\Phi_0$ represents the $O(\lambda^2)$ dissipative winding pump fueled by the bath phase current, $\chi_0 = \chi_0' + i\chi_0'' \in \mathbb{C}$ is the retarded phase-winding susceptibility governing $O(\lambda^2)$ coherent bath recoil, and $\Phi_1$ represents the leading $O(\lambda^4)$ dissipative back-reaction of the environment as the system accumulates phase-space volume.

Furthermore,  $e^{i\hat{x}(\tau)} := U_E^\dagger(\tau) e^{i\hat{x}} U_E(\tau)$ denotes the Heisenberg-picture phase-shift operator. In the Fourier basis $\{|n\rangle\}_{n \in \mathbb{Z}}$, the unequal finite-time commutator evaluates to $$[e^{i\hat{x}(\tau)}, e^{-i\hat{x}(0)}]|n\rangle = (\alpha_n^*(\tau)\alpha_{n-1}(\tau) - \alpha_{n+1}^*(\tau)\alpha_n(\tau))|n\rangle,$$ which is non-vanishing for $\tau > 0$ whenever the environmental oscillator possesses finite phase inertia or nonlinear dispersion ($E_{n+1} - 2E_n + E_{n-1} \neq 0$).

Each of the two subsystem nodes of the bipartite state $\rho(t)$ is characterized by a dual pair of lowering and raising transition operators $L_-^{(j)}$ and $L_+^{(j)} = (L_-^{(j)})^\dagger$ for $j \in \{1, 2\}$. Along the forward ($+$) and backward ($-$) branches of the Keldysh contour, projecting onto generalized coherent states $|\Omega_j^\pm(t)\rangle$ maps these transition operators onto complex path expectation values,
\begin{equation}
L_-^{(j)\pm}(t) = \langle \Omega_j^\pm(t)|L_-^{(j)}|\Omega_j^\pm(t)\rangle = \mathcal{R}_j^\pm(t) e^{-i\Phi_j^\pm(t)},
\end{equation}
where $\mathcal{R}_j^\pm(t)$ and $\Phi_j^\pm(t)$ are the local coherence radius and azimuthal phase of node $j$ along each contour branch. At the operator level, the macroscopic coherence amplitude of node $j$ is tracked by the expectation value $R_j(t) = |\mathrm{Tr}(\rho(t)L_-^{(j)})|$.

\subsection{The 2-Form Action principle and  fluctuation filtering}

Standard path-integral techniques (such as the Caldeira–Leggett model) rely on a classical action which is a 1-form integrated along one-dimensional trajectories i.e. lines. By Liouville's theorem, 1-form actions generate strictly self-adjoint fluctuation operators that conserve phase-space volume, rendering them topologically and structurally incapable of describing autonomous phase-space contraction onto a stable periodic orbit.

To incorporate non-conservative attractor dynamics natively (at the path level), we define the action \(S\) as the surface integral of a Poincaré-Cartan 2-form \(\Omega \) over a two-dimensional \(\epsilon \)-filament surface \(\Sigma(q)\) compactified around the reference trajectory \(\gamma(t)\) in extended phase space,
$$S=\int _{\Sigma (q)}\Omega ,\qquad \Omega =dp_{i}\land dq^{i}-dT\land dt-F_{i}(q,p)\,dq^{i}\land dt$$

where \((q^i, p_i)\) are the canonical phase-space coordinates of the environment, \(T(q,p)\) is the Kinetic energy, and $F_i(q,p)$ is the non-conservative force field generating the trajectory, and is responsible for phase-space expansion or contraction when $\nabla \cdot F \neq 0$.

Taking the second order functional variation of \(S\) with respect to transverse path deviations \(\xi^i(t)\) around the stationary orbit \(\gamma(t)\) yields a fluctuation quadratic form governed by a Sturm–Liouville (Gelfand) fluctuation operator,

$$G_{ij}=-\delta _{ij}\frac{d^{2}}{dt^{2}}+\Gamma _{ij}(t)\frac{d}{dt}+K_{ij}(t)$$
where \(K_{ij}(t)\) is the coordinate restoration stiffness matrix and \(\Gamma_{ij}(t)\) is the antisymmetric velocity-dependent friction matrix evaluated along the trajectory,

$$\Gamma _{ij}(t)=\left.\left(\frac{\partial \mathcal{F}_{i}}{\partial p_{j}}-\frac{\partial \mathcal{F}_{j}}{\partial p_{i}}\right)\right|{}_{\gamma (t)}$$

Because the first-derivative friction term breaks time-reversal symmetry (\(G_{ij} \neq G_{ij}^\dagger\)), evaluating the Gaussian path integral over the fluctuation histories via the Gelfand–Yaglom method and applying Abel’s Wronskian identity factors the functional determinant into a conservative oscillatory baseline \(\mathcal{D}_{\text{osc}}(T)^{-1/2}\) multiplied by a real exponential damping envelope,

$$\big[\det (G)\big]^{-1/2}\propto \big[\mathcal{D}_{\text{osc}}(T)\big]^{-1/2}\exp \!\left[-\frac{1}{4}\int _{0}^{T}\text{Tr}\big(\partial F_i/\partial p_i\big)\,dt\right]$$

Whenever the attractor satisfies the dissipative Lyapunov condition \(\text{Tr}(\partial F_i/\partial p_i) < 0\), this determinant prefactor acts as a geometric amplitude-focusing lens. It exponentially penalizes unstable fluctuation histories, physically concentrating the quantum wave packet directly onto the periodic attractor shell.

\subsection{Topological Sector Collapse and the Normalization Shield}

Because the support of \(\rho_E(t)\) is the compact circle manifold \(U(1)\), time-slicing the Keldysh contour into \(M\) intervals of duration \(\epsilon = T/M\) yields the short-time phase-bridge transition weight at step \(k\),

$$W_k^{\pm}(x_k) = \exp\!\left[ \mp 2i\epsilon\lambda \sum_{j=1}^2 \mathcal{R}_j^{\pm}(t_k) \cos\!\big(x_k - \Phi_j^{\pm}(t_k)\big) \right]$$

Expanding this periodic weight via the Jacobi–Anger identity (\(e^{z\cos\theta} = \sum_{m \in \mathbb{Z}} I_m(z) e^{im\theta}\)) decomposes the phase interaction into discrete integer winding harmonics \(m_k \in \mathbb{Z}\),

$$W_k^+(x_k) = \prod_{j=1}^{2} \sum_{m_{k,j} \in \mathbb{Z}} (-i)^{m_{k,j}} I_{m_{k,j}}\!\left(2\epsilon\lambda\mathcal{R}_j^+(t_k)\right) e^{im_{k,j}(x_k - \Phi_j^+(t_k))}$$

Integrating over the intermediate compact phase variables \(x_k \in [0, 2\pi)\) makes the Fourier basis states overlaps, 

$$\langle n_k \vert{} e^{im_k x} \vert{} n_{k-1} \rangle = \delta_{n_k, n_{k-1}+m_k}$$

This is the Topological Sector Collapse, continuous spatial path integration is replaced exactly by discrete integer jumps in the environmental winding number,
$$n_{k}=n_{k-1}+m_{k},\qquad \tilde{n}_{k}=\tilde{n}_{k-1}+\tilde{m}_{k}\qquad (m_{k},\tilde{m}_{k}\in \mathbb{Z})$$

At the final time \(t_M = T\), tracing out the environment imposes the boundary identity \(n_M = \tilde{n}_M = n\), locking in the total (net) winding number of the forward and backward Keldysh branches. In the continuum limit \(\epsilon \to 0\), summing over the intermediate Bessel weights compiles the boundary phase terms into the normalization prefactor, i.e. a Normalization Shield,

$$\mathcal{N}_{\text{shield}}=\exp \!\left[-2\lambda \int _{0}^{T}dt\sum _{j=1}^{2}\Big(\mathcal{R}_{j}^{+}(t)\cos \!\big(\Phi _{j}^{+}(t)-\Psi _{E}(t)\big)-\mathcal{R}_{j}^{-}(t)\cos \!\big(\Phi _{j}^{-}(t)-\Psi _{E}(t)\big)\Big)\right]$$

Along any (physical) diagonal configuration where the forward and backward paths coincide (\(\mathcal{R}_j^+ = \mathcal{R}_j^-\) and \(\Phi_j^+ = \Phi_j^-\)), the exponent vanishes identically, giving \(\mathcal{N}_{\text{shield}}\vert_{\text{diag}} = 1\). This guarantees that the path integral identically preserves the trace of the state $$\mbox{Tr}[{\dot{\rho} }(t)] = 0$$

\subsection{The Fundamental Bipartite Master Equation}

The continuous time derivative of the log influence functional $\frac{d}{dt}\ln\mathcal{F}$ evaluated at the upper limit $t$ maps the system's trajectory directly onto the reduced density matrix. Because each single-quantum transition operator $L_+^{(j)}$ exchanges exactly one topological winding quantum ($m_k = \pm 1$) with the $U(1)$ phase bridge, the leading environmental pumping functional $\Phi_0[\rho_E(t)]$ is dressed by the local phase-space volume recoil of the Normalization Shield, $-\Phi_1[\rho_E(t)] R_j^2(t)$.

Simultaneously, at order $O(\lambda)$, the system's transition current induces a retarded linear-response recoil $\delta Z(t)$ in the environmental order parameter $Z(t) = Z_0(t) + \delta Z(t)$. Evaluating the path variables at their state expectation values under the bath-factorized closure (Section~\ref{sec:defense}) yields, in the Markovian limit,
\begin{equation}
\delta Z(t) \approx \lambda \chi_0 \sum_{k=1}^2 \mathrm{Tr}\!\left(\rho(t) L_+^{(k)}\right).
\end{equation}
Substituting $Z(t) = Z_0(t) + \delta Z(t)$ into the phase-bridge interaction $H_{\mathrm{int}} = \hbar\lambda\sum_{j=1}^2(Z(t)L_-^{(j)} + Z^*(t)L_+^{(j)})$ separates the coherent generator into the bare entrainment Hamiltonian $K_{\mathrm{eff}}(t)$ and a state-dependent coherent recoil operator $\Delta(\rho)$:
\begin{equation}
K_{\mathrm{eff}}(t) = H_{\mathrm{bi}} + \underbrace{\hbar\lambda \sum_{j=1}^2 \left(Z_0(t) L_-^{(j)} + Z_0^*(t) L_+^{(j)}\right)}_{H_{\mathrm{ent}}(t)},
\label{eq:phaseInt}
\end{equation}
\begin{equation}
\Delta(\rho) = \sum_{j,k=1}^2 \left(g\,\mathrm{Tr}\!\left(\rho L_-^{(k)}\right) L_+^{(j)} + g^*\,\mathrm{Tr}\!\left(\rho L_+^{(k)}\right) L_-^{(j)}\right), \qquad g := \hbar\lambda^2 \chi_0^* = \alpha - i\beta \in \mathbb{C},
\label{eq:coherent_recoil}
\end{equation}
where $H_{\mathrm{bi}}$ is the internal bipartite Hamiltonian.\footnote{We use a bipartite system to study the indirect coupling and mutual phase locking of two subsystems via their shared interaction with a synchronizing phase oscillator.}

Collecting the coherent recoil $\Delta(\rho)$, the topological pump and Normalization Shield saturation, the linear relaxation rate $\Gamma_2$, the inter-node dissipative cross-talk block $J_{\mathrm{chiral}}[\rho]$, and local thermal channels $\mathcal{L}_{T,j}$ yields the fundamental master equation for the limit-cycle attractor:
\begin{equation}
\frac{d\rho}{dt} = -i\big[K_{\mathrm{eff}}(t) + \Delta(\rho),\, \rho\big] + \sum_{j=1}^2 \Big(\Phi_0[\rho_E(t)] - \Phi_1[\rho_E(t)]\big|\mathrm{Tr}\big(\rho L_-^{(j)}\big)\big|^2\Big)\mathcal{D}_{L_+^{(j)}}[\rho] + \sum_{j=1}^2 \Gamma_2 \mathcal{D}_{L_-^{(j)}}[\rho] + J_{\mathrm{chiral}}[\rho] + \sum_{j=1}^2 \mathcal{L}_{T,j}[\rho],
\label{eq:bipartite_master}
\end{equation}
where $\mathcal{D}_A[\rho] = A\rho A^\dagger - \frac{1}{2}\{A^\dagger A, \rho\}$ is the standard CPTP Lindblad superoperator, and $J_{\mathrm{chiral}}[\rho]$ captures the dissipative cross-node phase modulation,
\begin{equation*}
J_{\mathrm{chiral}}[\rho] = \chi \sum_{j \neq k}^2 \mathrm{Tr}\!\left(\rho L_-^{(k)}\right)\left(L_+^{(j)}\rho L_-^{(j)} - \frac{1}{2}\left\{L_-^{(j)}L_+^{(j)}, \rho\right\}\right) + \mathrm{h.c.} = 2\chi \sum_{j \neq k}^2 \mathrm{Re}\!\left[\mathrm{Tr}\!\left(\rho L_-^{(k)}\right)\right]\mathcal{D}_{L_+^{(j)}}[\rho],
\end{equation*}
with $\chi = \lambda^2 R_E(t)$.

Here, $\Delta(\rho)$ serves as the exact $O(\lambda^2)$ coherent counterpart to $J_{\mathrm{chiral}}[\rho]$ (with any interaction-picture transformation taken with respect to the bare Hamiltonian $H_{\mathrm{bi}}$). Both terms depend solely on the local reduced expectation values $\mathrm{Tr}(\rho L_-^{(k)})$ and arise from the same bath-mediated back-action. When the topological pump exceeds linear damping ($\Phi_0 > \Gamma_2$), the trivial focus at $R_j = 0$ loses stability: in infinite-dimensional bosonic systems, dissipative pumping drives $R_j$ outward until gain saturation locks it onto $R_{j,*}^2 = (\Phi_0 - \Gamma_2)/\Phi_1$; in finite-dimensional spin or qudit systems, $\Phi_0 > \Gamma_2$ establishes population inversion, which $\Delta(\rho)$ ($\beta > 0$) converts into coherent radial growth until population depletion stabilizes the limit cycle (Section~\ref{sec:correspondence}).

\subsubsection{Distinguishing the Topological Phase Bridge from Linear Jaynes--Cummings Dynamics}

The similarity of $H_{\mathrm{ent}}(t)$ in Eq.~(\ref{eq:phaseInt}) with single-photon driving suggests a superficial comparison with semiclassical linear drives such as the Jaynes--Cummings or Maxwell--Bloch models. Despite the formal algebraic resemblance, its physical origin and mathematical properties are fundamentally distinct due to the $U(1)$ limit-cycle topology of the environment:
\begin{enumerate}
\item \textbf{Non-Local Phase Entrainment vs.\ Field Subtraction:} In a linear coupling scheme, the system exchanges quanta with a passive reservoir, leading to contractive relaxation. Here, $Z_0(t) = \mathrm{Tr}_E(\rho_E(t)e^{ix})$ is the complex order parameter of a macroscopically synchronized phase condensate; the bridge mediates geometric phase locking of the system's transition currents onto the circular tracks of the environmental limit cycle.
\item \textbf{Nonlinear Feedback Duality:} A linear drive cannot alter the contractive nature of dissipative channels. With the phase-bridge coupling, the same Keldysh interaction that produces $H_{\mathrm{ent}}(t)$ at $O(\lambda)$ simultaneously generates both the coherent recoil $\Delta(\rho)$ at $O(\lambda^2)$ and the dissipative gain saturation $-\Phi_1|\mathrm{Tr}(\rho L_-^{(j)})|^2\mathcal{D}_{L_+^{(j)}}[\rho]$ at $O(\lambda^4)$, because the forward and backward Keldysh branches are bound by the global winding trace constraint $n_M = \tilde{n}_M$.
\item \textbf{Destabilization of the Vacuum Focus:} Unlike the undriven Jaynes--Cummings model, which relaxes to a stable ground state, the topological winding pump $\Phi_0 > \Gamma_2$ renders the origin $R_j = 0$ unstable, inducing the birth of an autonomous quantum limit cycle even in the uncoupled limit $Z_0 \to 0$.
\end{enumerate}

\subsection{Positivity Basin and Infinite-Dimensional Nagumo Invariance}

Because $\Delta(\rho)$ in Eq.~(\ref{eq:coherent_recoil}) is strictly Hermitian for every state $\rho$, the coherent commutator $-i[K_{\mathrm{eff}}(t) + \Delta(\rho), \rho]$ preserves trace normalization and density-matrix positivity globally on the entire state space. Positivity of the full generator therefore reduces to non-negativity of the dissipative pumping rates $\gamma_j(\rho) = \Phi_0 - \Phi_1|\mathrm{Tr}(\rho L_-^{(j)})|^2 + 2\chi\sum_{k \neq j}\mathrm{Re}[\mathrm{Tr}(\rho L_-^{(k)})]$.

In finite-dimensional systems where population depletion saturates the limit cycle ($\Phi_1 = 0$, $\chi = 0$), $\gamma_j = \Phi_0 > 0$ is constant and positivity holds globally on all of $\mathcal{S}(\mathcal{H}_{\mathrm{bi}})$ without restricting the state space. When $\Phi_1 > 0$ (as required for purely dissipative saturation in infinite dimensions), we restrict initial states to the closed, convex positive-rate basin
\begin{equation}
\mathcal{B} = \left\{ \rho \in \mathcal{S}(\mathcal{H}_{\mathrm{bi}}) \;\middle|\; \Phi_1 R_j^2(\rho) - 2\chi\sum_{k \neq j}\mathrm{Re}\!\left[\mathrm{Tr}\!\left(\rho L_-^{(k)}\right)\right] \leq \Phi_0, \;\; j \in \{1, 2\} \right\}.
\label{eq:positivity_basin}
\end{equation}
Notice that $\mathcal{B}$ is convex because $R_j^2(\rho)$ is convex and $\mathrm{Re}[\mathrm{Tr}(\rho L_-^{(k)})]$ is linear in $\rho$, and it reduces to $R_j^2(\rho) \leq \Phi_0/\Phi_1$ when $\chi = 0$ or in the collective charge-sector basis. On the boundary $\partial\mathcal{B}$ where equality holds, the pumping rate vanishes identically ($\gamma_j|_{\partial\mathcal{B}} = 0$), leaving only the coherent evolution and the contractive damping channels $\Gamma_2\mathcal{D}_{L_-^{(j)}}[\rho] + \mathcal{L}_{T,j}[\rho]$. Consequently, the radial drift velocity points strictly inward along the boundary,
\begin{equation}
\left.\frac{d}{dt}R_j^2\right|_{\partial\mathcal{B}} < 0,
\end{equation}
provided the cross-talk and entrainment couplings satisfy $2\chi\sum_{k \neq j} R_k < \Phi_0$ and $2\lambda R_E < \Gamma_2\sqrt{\Phi_0/\Phi_1}$ (ensuring linear contraction dominates on $\partial\mathcal{B}$).

Because $\gamma_j(\rho_0) \geq 0$ holds at every $\rho_0 \in \mathcal{B}$, freezing the state-dependent coefficients at any instant yields a valid linear GKSL superoperator admitting a standard Stinespring dilation. Using the generalized Lie--Trotter product formula (Appendix~\ref{app:domain_regularity}, Lemma~\ref{lem:trotter_addition}) to combine this generator with standard linear Lindblad channels yields a globally well-defined, trace-preserving, and positivity-preserving evolution on $\mathcal{B}$.

\section{Inverse Coupling Theorem}
\label{sec:inverse_topology}

We establish an inferred mapping backward from macroscopic kinematic behavior to its microscopic interaction. The persistent existence of a stable, entrainable quantum limit cycle acts as a structural constraint that strictly dictates the topological invariants of the background environment. We formalize this geometric constraint through the following theorem, which defines the boundary between dissipative relaxation and autonomous quantum limit cycles:

\begin{theorem}[The Inverse Coupling Theorem]
\label{thm:inverse_topology}
An autonomous, entrainable quantum limit cycle\footnote{Corresponding to a supercritical Hopf or SNIPER bifurcation in the classical limit. The manifold $\mathcal{M}$ here refers to the limit set of the background phase oscillator(s) rather than its full ambient phase space.} occurs in the reduced dynamics of an open quantum system ($\dim\mathcal{H} \leq \aleph_0$) if and only if the underlying system--environment interaction manifold $\mathcal{M}$ possesses a non-trivial first homotopy group $\pi_1(\mathcal{M}) = \mathbb{Z}$ and the topological winding pump exceeds linear damping ($\gamma_q^+(0) > \gamma_q^-(0)$, i.e., $\Phi_0 > \Gamma_2$). This selects the compact circle group $S^1 \cong U(1)$ as the unique valid interaction topology capable of sustaining periodic attractors.

Conversely, simply connected interaction manifolds ($\pi_1(\mathcal{M}) = 0$) extinguish the topological winding pump ($\Phi_0 = 0$), forcing the dynamics unconditionally onto stationary fixed points in both finite and infinite dimensions.
\end{theorem}

To prove this result for a generic density matrix $\rho(t)$ of arbitrary dimension ($\dim\mathcal{H} \leq \aleph_0$), we analyze the mapping between the functional integration of the 2-form action $S = \int_{\Sigma}\Omega$ and the stability of the resulting reduced generator $\mathcal{L}$. For a quantum state to exhibit a phase-coherent limit cycle, its linearized dynamics around the periodic orbit must possess an isolated pair of purely imaginary conjugate Floquet/orbital frequencies $\pm i\Omega_0$ ($\Omega_0 > 0$) while all transverse radial modes retain strictly negative real parts ($\mathrm{Re}\,\lambda_{\mathrm{rad}} < 0$), without requiring external driving ($Z_0 = 0$).

\subsection{Exact Winding Selection and Harmonic Series Truncation}

Consider an arbitrary periodic interaction potential $\mathcal{V}(x)$ on a compact domain $x \in [0, 2\pi)$. Expanding $\mathcal{V}(x)$ into its general Fourier series yields higher-harmonic components,
\begin{equation}
\mathcal{V}(x) = \sum_{k=1}^{\infty}\Big[g_k \sin(kx) + h_k \cos(kx)\Big] = \sum_{k=1}^{\infty} \chi_k \cos\big(k(x - \phi_k)\big),
\end{equation}
where $\chi_k = \sqrt{g_k^2 + h_k^2}$ and $\phi_k$ are the amplitude and phase offset of the $k$-th harmonic mode. Along the Keldysh contour, each $k$-th harmonic activates a multi-winding transition weight expanded via the Jacobi--Anger identity,
\begin{equation}
\exp\!\left[-2i\epsilon\lambda \sum_{j=1}^2 \mathcal{R}_j^+(t)\mathcal{V}(x)\right] = \prod_{k=1}^{\infty} \sum_{m_k \in \mathbb{Z}} (-i)^{m_k} I_{m_k}\!\big(2\epsilon\lambda\chi_k \mathcal{R}_j^+(t)\big) e^{im_k k(x - \phi_k)}.
\end{equation}
Integrating over the intermediate compact phase coordinate $x \in [0, 2\pi)$ enforces exact integer angular-momentum conservation across the environmental Fourier basis $\{|n\rangle\}_{n \in \mathbb{Z}}$, $\Delta n = \sum_{k=1}^{\infty} k m_k$. This topological selection rule strictly classifies every harmonic mode $k \geq 2$ into one of two exact algebraic outcomes, completely independent of the coupling strength $\lambda$:
\begin{enumerate}
\item \textbf{Exact Decoupling of Multi-Winding Jumps ($k m_k \neq \pm 1$):} Any single-quantum system transition operator $L_+^{(j)}$ carries fundamental charge $\Delta n = \pm 1$. Because $k \geq 2$ implies $k m_k \in \{0, \pm k, \pm 2k, \dots\}$, no non-zero winding jump $m_k \neq 0$ of a higher harmonic can match the $\Delta n = \pm 1$ selection rule of $L_\pm^{(j)}$. Higher harmonics couple exclusively to multi-quantum transition operators $(L_+^{(j)})^k$ (which vanish identically by nilpotency in $d_j = 2$ qubits and decouple orthogonally from the single-quantum channel in arbitrary dimensions).
\item \textbf{Absorption into Existing Linear Channels ($m_k = 0$):} The zero-winding sector ($m_k = 0$) of every higher harmonic $k \geq 2$ carries topological charge $\Delta n = 0$. Its trace contribution yields phase-insensitive scalar rates that are absorbed identically into the linear damping and thermal channels ($\Gamma_2$ and $\mathcal{L}_{T,j}$) without altering the functional form of the generator.
\end{enumerate}
Thus, the fundamental $k = 1$ sinusoidal harmonic is selected non-perturbatively as the unique active phase-bridge interaction within the single-quantum sector.

\subsection{Topological Obstruction for Simply Connected Manifolds ($\pi_1(\mathcal{M}) = 0$) and Uniqueness of $S^1$}

To establish the reverse implication ($\text{Limit Cycle} \implies \pi_1(\mathcal{M}) \cong \mathbb{Z}$), suppose that the reduced dynamics is realized via a path integral over a connected, smooth background interaction manifold $\mathcal{M}$.

\begin{itemize}
\item \textbf{Step 1 (Keldysh Loop Homotopy and Vanishing of Winding Pumping):} Along the closed-time Keldysh contour $[0, T]$, the forward trajectory $x^+(t) \in \mathcal{M}$ and backward trajectory $x^-(t) \in \mathcal{M}$ originate from the same environmental support and are identified at the final trace boundary $x^+(T) = x^-(T)$. Their concatenation $\gamma_{\mathrm{K}} := x^+ \circ (x^-)^{-1}$ defines a closed loop in $\mathcal{M}$ and hence a homotopy class $[\gamma_{\mathrm{K}}] \in \pi_1(\mathcal{M})$. Suppose for contradiction that $\mathcal{M}$ is simply connected, $\pi_1(\mathcal{M}) = 0$ (which includes all contractible non-compact spaces $\mathcal{M} \simeq \mathbb{R}^N$, as in the Caldeira--Leggett oscillator bath). Then every closed Keldysh loop $\gamma_{\mathrm{K}}$ is null-homotopic ($[\gamma_{\mathrm{K}}] = 0$). By the monodromy theorem, any flat $U(1)$ phase-bridge connection $\omega$ on $\mathcal{M}$ is globally exact ($\omega = d\theta$ for a single-valued real function $\theta : \mathcal{M} \to \mathbb{R}$) rather than circle-valued. Consequently, the environmental Hilbert space admits no discrete integer Fourier winding basis $\{|n\rangle\}_{n \in \mathbb{Z}}$ and no quantized winding jumps ($n_k \in \mathbb{Z}$), regardless of whether the system--bath coupling is linear or an arbitrary smooth nonlinear function on $\mathcal{M}$. Without non-trivial winding currents around closed Keldysh loops, the topological winding pump vanishes identically: $\Phi_0[\rho_E] = 0$ (and $\Phi_1[\rho_E] = 0$).

\item \textbf{Step 2 (Collapse to Stationary Fixed Points in Both Finite and Infinite Dimensions):} When $\Phi_0 = 0$, the upward pumping channel drops out of the reduced master equation ($\gamma_q^+(0) = 0$).
\begin{itemize}
\item In \emph{infinite dimensions} ($\dim\mathcal{H} = \aleph_0$, $\Delta(\rho) = 0$), the generator collapses to a strictly state-independent linear GKSL semigroup $\mathcal{L}_{\mathrm{lin}}$ with $\Gamma_2 > 0$. By the Kossakowski--Lindblad contraction theorem, $\frac{d}{dt}\|\rho_1(t) - \rho_2(t)\|_1 \leq 0$, forbidding isolated purely imaginary eigenvalues $\pm i\Omega_0$ and pulling all trajectories onto a stationary fixed point.
\item In \emph{finite dimensions} ($\dim\mathcal{H} < \infty$), even if a passive damped bath mode still generates a retarded coherent recoil $\Delta(\rho)$ with $\beta = \hbar\lambda^2 \chi_0'' > 0$, the absence of upward pumping ($\Phi_0 = 0$) forces the system out of population inversion ($\langle [L_+, L_-]\rangle < 0$, e.g., $\langle \sigma_z \rangle \to -1$). Consequently, the coherent recoil contributes strictly negative radial drift ($\beta \langle \sigma_z \rangle < 0$, acting as Purcell damping), collapsing every trajectory unconditionally to the stationary ground state.
\end{itemize}
Thus $\pi_1(\mathcal{M}) = 0$ strictly forbids an autonomous quantum limit cycle in all dimensions.

\item \textbf{Step 3 (Abelianization and Reduction to $\pi_1(\mathcal{M}) \cong \mathbb{Z}$):} By Steps~1 and~2, sustaining an autonomous limit cycle requires $\pi_1(\mathcal{M}) \neq 0$. By Axiom~(P1), the system couples via integer $U(1)$ phase charges $\Delta n = \pm q$, so environmental holonomies enter the Keldysh influence functional exclusively through unitary characters $\chi \in \mathrm{Hom}(\pi_1(\mathcal{M}), U(1))$. Because $U(1)$ is abelian, every such character annihilates all commutators and factors through the abelianization $\pi_1(\mathcal{M})^{\mathrm{ab}}$. Finite torsion subgroups yield only discrete roots of unity rather than continuous $U(1)$ equivariance, whereas multiple independent cyclic factors ($\mathbb{Z}^k$, $k \geq 2$) span a higher-dimensional quasi-periodic torus $T^k$. Projecting onto a single periodic limit-cycle orbit $\mathcal{M}_{\mathrm{att}} \cong S^1$---together with the orthogonal decoupling of higher harmonics $k \geq 2$---selects a single infinite cyclic winding factor $\pi_1(\mathcal{M}_{\mathrm{eff}}) \cong \mathbb{Z}$, uniquely identifying the effective interaction manifold with $S^1 \cong U(1)$.
\end{itemize}

\textit{Sufficiency ($\Leftarrow$).} Conversely, if $\pi_1(\mathcal{M}) \cong \mathbb{Z}$ so that $\mathcal{M}_{\mathrm{eff}} \cong S^1 \cong U(1)$ with active phase-bridge pumping $\Phi_0 > \Gamma_2$, the two-form Keldysh path integral of Section~\ref{sec:setup} yields Eq.~(\ref{eq:bipartite_master}). In infinite dimensions ($\Phi_1 > 0$), the origin $R_j = 0$ is linearly unstable while the outer boundary $R_j^2 = \Phi_0/\Phi_1$ is strictly inward-pointing, yielding an isolated limit cycle at $R_{j,*} = \sqrt{(\Phi_0 - \Gamma_2)/\Phi_1} > 0$. In finite dimensions ($\beta > 0$, $\Phi_1 = 0$), $\Phi_0 > \Gamma_2$ establishes population inversion; above the Hopf threshold $2\beta(\Phi_0 - \Gamma_2) > (\Phi_0 + \Gamma_2)^2$, $\Delta(\rho)$ amplifies coherence until population depletion locks the state onto a unique hyperbolic limit cycle inside the Bloch ball (Section~\ref{sec:correspondence}). \qed

\paragraph*{Exclusion of non-compact topologies ($\pi_1(\mathcal{M}) = 0$)}
Whether the coupling to a non-compact coordinate $q \in \mathbb{R}$ is linear or an arbitrary smooth nonlinear function $\mathcal{V}(Q_j, q)$ on $\mathcal{M} = \mathbb{R}$, the contractibility of the manifold ($\pi_1(\mathbb{R}) = 0$) precludes discrete topological winding sectors ($n_k \in \mathbb{Z}$) and bounds the dynamics unconditionally to stationary states.

\section{Unified Structure Theorem for Quantum Phase Oscillators}
\label{sec:structure}

Demonstrating the topological necessity of the $U(1)$ phase bridge allows us to unify our geometric and operator-algebraic results into a single dimension-independent classification theorem. Crucially, our result holds across arbitrary separable Hilbert spaces---both finite-dimensional spins/qudits ($\dim\mathcal{H} < \infty$) and countably infinite-dimensional oscillator spaces ($\dim\mathcal{H} = \aleph_0$)-without assuming a polynomial truncation upfront.

\subsection{Axiomatic Class of Entrainable Phase Oscillators}

Let $\mathcal{H}$ be a separable Hilbert space ($\dim\mathcal{H} \leq \aleph_0$), let $\mathfrak{T}_{1,2}(\mathcal{H})$ denote the weighted space of finite-energy trace-class states, and let $\mathcal{D}_0 \subset \mathfrak{T}_{1,2}(\mathcal{H})$ be the dense finite-level core defined in Appendix~\ref{app:domain_regularity}. We define the class $\mathfrak{G}_{\mathrm{osc}}$ of entrainable quantum phase-oscillator generators as all continuous, trace-preserving dynamical generators $\mathcal{L}_{Z_0} : \mathfrak{T}_{1,2}(\mathcal{H}) \to \mathfrak{T}_1(\mathcal{H})$ parameterized by an environmental order parameter $Z_0(t) = R_E e^{i\Omega_0 t}$ that satisfy three physical axioms:

\begin{description}
\item[(P1) Joint $U(1)$ Equivariance] The system carries a strongly continuous unitary representation of the circle group $U_\theta = e^{i\theta Q}$ ($\theta \in [0, 2\pi)$) generated by a self-adjoint phase-charge operator $Q$ with discrete integer spectrum $\sigma(Q) \subset \mathbb{Z}$. The generator is jointly equivariant under simultaneous phase rotations of the system and environment:
\begin{equation}
\mathcal{L}_{e^{-i\theta}Z_0}\big(U_\theta \rho U_\theta^\dagger\big) = U_\theta \mathcal{L}_{Z_0}(\rho) U_\theta^\dagger \qquad \forall \theta \in [0, 2\pi),
\end{equation}
which reduces to global $U(1)$ covariance $\mathcal{L}_0(U_\theta \rho U_\theta^\dagger) = U_\theta \mathcal{L}_0(\rho) U_\theta^\dagger$ in the uncoupled limit $Z_0 = 0$.

\item[(P2) Symmetric Bidirectional Phase Interface] Within each active integer charge sector $q \geq 1$, the system and the $U(1)$ background oscillator interact exclusively through a single dual pair of charge-$\pm q$ transition currents ($L_\pm^{(q)}$ on the system, $e^{\pm iqx}$ on the bath), such that the reduced generator depends smoothly on the system state $\rho$ only through the dual order parameter $z_q(\rho) = \mathrm{Tr}(\rho L_-^{(q)})$, while every frozen-state generator $\mathcal{L}_{Z_0,\,\rho_0}[\cdot]$ on a convex invariant basin $\mathcal{B}_{1,2} \cap \mathcal{D}_0$ lies within the completely positive GKSL Lindblad cone.

\item[(P3) Entrainable Hyperbolic $S^1$ Limit Cycle] In the uncoupled limit $Z_0 = 0$, trajectories inside $\mathcal{B}_{1,2}$ (away from the measure-zero phase-symmetric subspace $z_q = 0$) converge to an isolated, structurally stable (hyperbolic) periodic orbit $\mathcal{M}_{\mathrm{att}} \cong S^1$ of finite coherence radius $R_{q,*} > 0$ with strictly negative radial contraction rate ($\mathrm{Re}\,\lambda_{\mathrm{rad}} < 0$) and orbital frequencies $\pm i q \Omega_0$. For weak non-zero $Z_0(t) \neq 0$, this orbit phase-locks (entrains) to the environmental phase.
\end{description}

Requiring the limit cycle to persist at $Z_0 = 0$ in Axiom~(P3) cleanly distinguishes genuine quantum entrainment from passive forced oscillation.

\subsection{Statement and Proof of the Structure Theorem}

\begin{theorem}[Structure Theorem for Entrainable Quantum Phase Oscillators]
\label{thm:structure}
Let $\mathcal{H}$ be an arbitrary separable Hilbert space ($\dim\mathcal{H} \leq \aleph_0$), and let $\mathcal{L}_{Z_0} \in \mathfrak{G}_{\mathrm{osc}}$ satisfy Axioms~(P1)--(P3). Then $\mathcal{L}_{Z_0}$ generates a globally well-defined, trace-preserving, positivity-preserving entrainable limit cycle on $\mathcal{B}_{1,2}$ if and only if:
\begin{enumerate}
\item[(I)] \textbf{Universal Operator Form:} Up to unitary equivalence and phase-covariant linear GKSL channels $\mathcal{L}_{\mathrm{lin}}$, the generator takes the exact form summed over active integer charge sectors $q \in \mathbb{N}$,
\begin{equation}
\mathcal{L}_{Z_0}(\rho) = -i\Big[H_0 + \sum_{q \geq 1} h_q(|z_q|^2)Q + \Delta(\rho) + H_{\mathrm{ent}}(t),\, \rho\Big] + \sum_{q \geq 1}\gamma_q^+(|z_q|^2)\mathcal{D}_{L_+^{(q)}}[\rho] + \sum_{q \geq 1}\gamma_q^-(|z_q|^2)\mathcal{D}_{L_-^{(q)}}[\rho] + \mathcal{L}_{\mathrm{lin}}[\rho],
\label{eq:universal_structure}
\end{equation}
where $[Q, H_0] = 0$, $[Q, L_\pm^{(q)}] = \pm q L_\pm^{(q)}$, $z_q = \mathrm{Tr}(\rho L_-^{(q)})$, $h_q : \mathbb{R}_{\geq 0} \to \mathbb{R}$, $g_q : \mathbb{R}_{\geq 0} \to \mathbb{C}$, and $\gamma_q^\pm : \mathbb{R}_{\geq 0} \to \mathbb{R}_{\geq 0}$ are smooth scalar functions on $\mathcal{B}_{1,2}$, with coherent recoil and entrainment terms
\begin{equation}
\Delta(\rho) = \sum_{q \geq 1}\Big(g_q(|z_q|^2) z_q L_+^{(q)} + g_q^*(|z_q|^2) z_q^* L_-^{(q)}\Big), \qquad H_{\mathrm{ent}}(t) = \hbar\lambda \sum_{q \geq 1}\Big(Z_0(t)^q L_-^{(q)} + Z_0^*(t)^q L_+^{(q)}\Big).
\end{equation}

\item[(II)] \textbf{Dimensional Dichotomy:} Density-matrix positivity splits the active limit-cycle mechanism into two distinct algebraic regimes:
\begin{itemize}
\item \emph{Finite Dimensions ($\dim\mathcal{H} < \infty$):} Purely dissipative gain saturation ($\Delta(\rho) = 0$) is strictly forbidden by trace-norm contraction of frozen-CPTP maps. Sustaining a limit cycle requires active coherent recoil ($\mathrm{Im}\,g_q(0) < 0$) paired with population-inverting pumping ($\gamma_q^+(0) > \gamma_q^-(0)$).
\item \emph{Infinite Dimensions ($\dim\mathcal{H} = \aleph_0$, $[L_-^{(q)}, L_+^{(q)}] > 0$):} Unbounded ladder operators bypass the trace-norm obstruction, permitting purely dissipative limit cycles ($\Delta(\rho) = 0$) whenever the net gain $\Delta\gamma_q(s) := \gamma_q^+(s) - \gamma_q^-(s)$ satisfies $\Delta\gamma_q(0) > 0$, $\Delta\gamma_q(R_{q,*}^2) = 0$, and $\Delta\gamma_q'(R_{q,*}^2) < 0$.
\end{itemize}

\item[(III)] \textbf{Topological Necessity:} In both regimes, destabilizing the origin requires $\gamma_q^+(0) > \gamma_q^-(0) > 0$. Any two-form path-integral realization of $\mathcal{L}_{Z_0}$ therefore acts over an interaction manifold $\mathcal{M}$ with $\pi_1(\mathcal{M}) = \mathbb{Z}$, uniquely selecting $S^1 \cong U(1)$.
\end{enumerate}
\end{theorem}

\begin{proof}
\textbf{($\Rightarrow$) Necessity.} Assume $\mathcal{L}_{Z_0} \in \mathfrak{G}_{\mathrm{osc}}$ satisfies Axioms~(P1)--(P3).

\emph{Step 1: Spectral Charge Decomposition and Equivariant Operator Form.} By the Peter--Weyl theorem for $U(1)$, $\mathcal{H} = \bigoplus_{n \in \mathbb{Z}} \mathcal{H}_n$ and operators decompose into eigen-operators $L_\pm^{(q)}$ satisfying $U_\theta L_\pm^{(q)} U_\theta^\dagger = e^{\pm iq\theta} L_\pm^{(q)}$, while $z_q(U_\theta \rho U_\theta^\dagger) = e^{-iq\theta} z_q(\rho)$. By Axiom~(P2), the generator is built exclusively from the neutral operators $(H_0, Q)$, the interface pair $L_\pm^{(q)}$, and smooth functions of $z_q$ and $Z_0$.
\begin{itemize}
\item \emph{Coherent part:} Under $U_\theta$, $Z_0^q L_-^{(q)}$ and $z_q L_+^{(q)}$ are strictly invariant, while $|z_q|^2$ is $U(1)$-invariant. Consequently, the most general Hermitian Hamiltonian consistent with Axioms~(P1)--(P2) is $H_0 + \sum_q h_q(|z_q|^2)Q + \Delta(\rho) + H_{\mathrm{ent}}(t)$.
\item \emph{Dissipative part:} Expanding any frozen-state GKSL dissipator in the interface pair $\{L_+^{(q)}, L_-^{(q)}\}$, cross-terms $\mathcal{D}_{L_+^{(q)}, L_-^{(q)}}$ transform with phase $e^{\pm 2iq\theta}$ and must vanish under $U(1)$ covariance at $Z_0 = 0$. The remaining diagonal channels $\mathcal{D}_{L_+^{(q)}}$ and $\mathcal{D}_{L_-^{(q)}}$ must carry $U(1)$-invariant non-negative scalar rates $\gamma_q^\pm(|z_q|^2) \geq 0$ on $\mathcal{B}_{1,2}$, yielding Eq.~(\ref{eq:universal_structure}).
\end{itemize}

\emph{Step 2: Proof of the Dimensional Dichotomy.} Set $Z_0 = 0$ and consider the autonomous dynamics.
\begin{itemize}
\item \emph{Finite-Dimensional No-Go for $\Delta(\rho) = 0$:} Suppose $\dim\mathcal{H} < \infty$ and $\Delta(\rho) = 0$. Then at every frozen state $\rho_0 \in \mathcal{B}_{1,2}$, the frozen linear GKSL generator $\mathcal{L}_{0,\,\rho_0}$ is itself $U(1)$-covariant and therefore commutes with the diagonal projection $\mathcal{P}_0(\sigma) := \frac{1}{2\pi}\int_0^{2\pi} U_\theta \sigma U_\theta^\dagger\,d\theta$. Consequently, the coherence component $\delta\rho(t) := \rho(t) - \mathcal{P}_0(\rho(t))$ obeys $\frac{d}{dt}\delta\rho = \mathcal{L}_{0,\,\rho(t)}[\delta\rho]$. Because $\mathcal{L}_{0,\,\rho(t)}$ generates a CPTP contraction at every instant, the upper right Dini derivative satisfies $\frac{d^+}{dt}\|\delta\rho(t)\|_1 \leq 0$. Since $\|L_-^{(q)}\|_{\mathrm{op}} < \infty$ in finite dimensions, $R_q(t) = |\mathrm{Tr}(\delta\rho(t) L_-^{(q)})| \leq \|L_-^{(q)}\|_{\mathrm{op}}\|\delta\rho(0)\|_1$, preventing any small perturbation from $R_q = 0$ from growing to a finite limit cycle $R_{q,*} > 0$. Thus $\Delta(\rho) \neq 0$ is mandatory in $\dim\mathcal{H} < \infty$; evaluating $\frac{d}{dt}z_q$ shows that $\Delta(\rho)$ contributes $(i\mathrm{Re}\,g_q - \mathrm{Im}\,g_q)\langle[L_+^{(q)}, L_-^{(q)}]\rangle z_q$, which amplifies coherence from $z_q = 0$ only when $\mathrm{Im}\,g_q(0) < 0$ and the system is population-inverted ($\gamma_q^+(0) > \gamma_q^-(0)$).
\item \emph{Infinite-Dimensional Dissipative Regime:} When $\dim\mathcal{H} = \aleph_0$ with $[L_-^{(q)}, L_+^{(q)}] = \tilde{\Delta}_q > 0$, $\|L_-^{(q)}\|_{\mathrm{op}} = \infty$ bypasses the finite-dimensional bound. Setting $\Delta(\rho) = 0$, the radial drift on a single active sector obeys $\frac{dR_q}{dt} = \frac{\delta_0}{2}\Delta\gamma_q(R_q^2) R_q$ (with $\delta_0 > 0$). Destabilizing $R_q = 0$ requires $\Delta\gamma_q(0) > 0$; an isolated orbit at $R_{q,*} > 0$ requires $\Delta\gamma_q(R_{q,*}^2) = 0$; and strict hyperbolic contraction $\mathrm{Re}\,\lambda_{\mathrm{rad}} = \delta_0 R_{q,*}^2 \Delta\gamma_q'(R_{q,*}^2) < 0$ requires $\Delta\gamma_q'(R_{q,*}^2) < 0$.
\end{itemize}

\emph{Step 3: Topological Necessity.} In both regimes, $\gamma_q^+(0) > \gamma_q^-(0) > 0$ is mandatory, so Theorem~\ref{thm:inverse_topology} forces $\pi_1(\mathcal{M}) = \mathbb{Z}$ and $\mathcal{M}_{\mathrm{eff}} \cong S^1 \cong U(1)$.

\medskip
\noindent\textbf{($\Leftarrow$) Sufficiency.} Given Eq.~(\ref{eq:universal_structure}) with the conditions in~(II)--(III), frozen-state Hermiticity of the coherent part and non-negativity of $\gamma_q^\pm(|z_q|^2)$ on $\mathcal{B}_{1,2}$ guarantee global existence, trace preservation, and positivity (globally on $\mathcal{S}(\mathcal{H})$ when $\gamma_q^\pm \geq 0$ everywhere, and via Nagumo invariance on $\mathcal{B}_{1,2}$ when $\gamma_q^+$ has a finite zero; Appendix~\ref{app:domain_regularity}). The conditions in~(II) guarantee a supercritical Hopf bifurcation from $R_q = 0$ to a unique hyperbolic $S^1$ limit cycle at $Z_0 = 0$ (on $\mathcal{B}_{1,2} \setminus \{R_q = 0\}$ in infinite dimensions, with the $R_q = 0$ subspace broken immediately for any $Z_0 \neq 0$), which entrains under weak $H_{\mathrm{ent}}(t)$. 
\end{proof}

\subsection{Leading-Order Keldysh Expansion and Perturbative Hierarchy}

For concrete computations, the analytic feedback functionals in Theorem~\ref{thm:structure} expand systematically in powers of the Keldysh coupling $\lambda^2$ via the $(2m+2)$-point connected bath winding cumulants:
\begin{equation}
g_q(s_q) = \sum_{m=0}^{\infty}\lambda^{2m+2} g_q^{(m)} s_q^m, \qquad \gamma_q^{\pm}(s_q) = \sum_{m=0}^{\infty}\lambda^{2m+2}\gamma_q^{\pm(m)} s_q^m, \qquad h_q(s_q) = \sum_{m=1}^{\infty}\lambda^{2m+2} h_q^{(m)} s_q^m,
\label{eq:keldysh_hierarchy}
\end{equation}
where $s_q := |z_q|^2$. Truncating at leading non-trivial order immediately recovers the canonical master equation of Section~\ref{sec:setup}:

\begin{corollary}[Leading-Order Master Equation]
\label{cor:keldysh_truncation}
Retaining the leading $O(\lambda^2)$ coherent recoil ($g_q = \hbar\lambda^2 \chi_{0,q}^*$, $h_q = 0$) and leading $O(\lambda^2, \lambda^4)$ dissipative cumulants ($\gamma_q^+ = \Phi_0^{(q)} - \Phi_1^{(q)}|z_q|^2$, $\gamma_q^- = \Gamma_2^{(q)}$) reduces Eq.~(\ref{eq:universal_structure}) to the canonical polynomial form:
\begin{equation}
\mathcal{L}_{Z_0}(\rho) = -i\big[H_0 + \Delta(\rho) + H_{\mathrm{ent}}(t),\, \rho\big] + \sum_{q \geq 1}\Big(\Phi_0^{(q)} - \Phi_1^{(q)}|z_q|^2\Big)\mathcal{D}_{L_+^{(q)}}[\rho] + \sum_{q \geq 1}\Gamma_2^{(q)}\mathcal{D}_{L_-^{(q)}}[\rho] + \mathcal{L}_{\mathrm{lin}}[\rho].
\label{eq:keldysh_corollary}
\end{equation}
\end{corollary}

\begin{remark}[Higher-Order Cumulants and Strong-Coupling Resummation]
The hierarchy in Eq.~(\ref{eq:keldysh_hierarchy}) provides a systematic prescription beyond standard supercritical Hopf limit cycles:
\begin{enumerate}
\item \emph{Subcritical and Bistable Regimes:} Retaining the $O(\lambda^6)$ term ($m=2$, proportional to $|z_q|^4$) and $h_q^{(1)}|z_q|^2 Q$ captures subcritical Hopf bifurcations (bistable limit cycles where $\Phi_1 < 0$ and $\Phi_2 > 0$) and nonlinear Kerr frequency pulling.
\item \emph{Pad\'e Resummation and Semiclassical Laser Saturation:} In strong-coupling bosonic regimes where the truncated polynomial $\Phi_0 - \Phi_1|z_q|^2$ approaches the Nagumo boundary $\partial\mathcal{B}_{1,2}$, Pad\'e resummation yields the globally positive saturation rate $\gamma_q^+(|z_q|^2) = \Phi_0 / [1 + (\Phi_1/\Phi_0)|z_q|^2]$, recovering standard semiclassical laser gain saturation and eliminating the need for a restricted positivity basin.
\end{enumerate}
\end{remark}

\section{Positivity Compatibility}
\label{sec:defense}

Nonlinear master equations are frequently critiqued for violating the standard axiomatic structure of Completely Positive and Trace-Preserving (CPTP) dynamical semigroups or introducing unphysical causal artifacts. Here, we demonstrate that the unified limit-cycle generator of Theorem~\ref{thm:structure} and Corollary~\ref{cor:keldysh_truncation} is a mathematically rigorous structure that preserves density-matrix positivity within an invariant physical domain and remains fully consistent with quantum causality.

\subsection{Positivity and Instantaneous Dynamical Structure}

A standard mathematical objection to master equations with state-dependent coefficients is that nonlinear maps do not generally satisfy linear complete positivity and can generate negative eigenvalues if an effective transition rate becomes negative. While the Ando--Choi structure theorem~\cite{Ando1986,Belavkin1986} characterizes globally completely positive nonlinear maps via infinite series of $(m,n)$-mixed homogeneous forms with positive-semidefinite block witnesses, state-dependent master equations with saturating gain require a trajectory-level positivity guarantee.

First, observe that the coherent recoil $\Delta(\rho)$ in Eq.~(\ref{eq:coherent_recoil}) is strictly Hermitian for every density matrix $\rho$. Freezing the state at any instant $\rho_0$ yields a standard unitary commutator $-i[K_{\mathrm{eff}}(t) + \Delta(\rho_0), \cdot]$, which preserves trace normalization and density-matrix positivity globally on the entire state space $\mathcal{S}(\mathcal{H})$.

Positivity of the full generator therefore depends solely on the non-negativity of the dissipative rate functionals $\gamma_q^\pm(|z_q|^2)$:
\begin{enumerate}
\item \textbf{Globally Positive Regime:} Whenever $\gamma_q^\pm(s) \geq 0$ for all $s \geq 0$---which includes finite-dimensional spin/qudit realizations with $\Phi_1 = 0$ (Section~\ref{sec:correspondence}) as well as Pad\'e-resummed laser saturation $\gamma_q^+(s) = \Phi_0/[1 + (\Phi_1/\Phi_0)s]$---no rate can ever turn negative, and positivity holds globally on all of $\mathcal{S}(\mathcal{H})$ without restricting the state space.
\item \textbf{Truncated Polynomial Regime ($\Phi_1 > 0$):} Isolating the leading-order gain-saturation channel of Corollary~\ref{cor:keldysh_truncation},
\begin{equation}
\mathcal{L}_{\mathrm{top}}(\rho) = \gamma(\rho)\mathcal{D}_{L_+}[\rho] = \Big(\Phi_0 - \Phi_1\big|\mathrm{Tr}(\rho L_-)\big|^2\Big)\mathcal{D}_{L_+}[\rho],
\end{equation}
multiplying $\Phi_0$ by the unit-trace identity $\mathrm{Tr}(\rho)\mathrm{Tr}(\rho^\dagger) = 1$ casts $\gamma(\rho)$ into a degree-$(1,1)$ sesquilinear trace form,
\begin{equation}
\gamma(\rho) = \Phi_0\mathrm{Tr}(\rho)\mathrm{Tr}(\rho^\dagger) - \Phi_1\mathrm{Tr}(\rho L_-)\mathrm{Tr}(\rho^\dagger L_+) = \mathrm{Tr}\big[(\rho \otimes \rho^\dagger)\,\Xi\big],
\end{equation}
with the self-adjoint two-copy structural witness $\Xi = \Phi_0(\mathbb{I} \otimes \mathbb{I}) - \Phi_1(L_- \otimes L_+^T)$ on $\mathcal{H} \otimes \mathcal{H}$. Because $\Phi_1 > 0$, $\Xi$ is not globally positive semidefinite; rather, $\gamma(\rho) \geq 0$ holds inside the closed, convex positive-rate basin
\begin{equation}
\mathcal{B}_{1,2} = \left\{\rho \in \mathfrak{T}_{1,2}(\mathcal{H}) : \mathrm{Tr}(\rho) = 1,\; \big|\mathrm{Tr}(\rho L_-)\big|^2 \leq \frac{\Phi_0}{\Phi_1}\right\}.
\end{equation}
As proved in Appendix~\ref{app:domain_regularity} (Lemmas~\ref{lem:energy_bound}--\ref{lem:tangent_cptp}), Nagumo's invariance theorem~\cite{Nagumo1942} guarantees that $\mathcal{B}_{1,2}$ is positively invariant under the semiflow, ensuring $\gamma(\rho(t)) \geq 0$ for all $t \geq 0$ whenever $\rho(0) \in \mathcal{B}_{1,2}$.
\end{enumerate}

In both regimes, freezing the state-dependent coefficients at any instant $t_0$ and physical state $\rho_0 = \rho(t_0)$ yields an exact linear GKSL superoperator $\mathcal{L}_{\rho_0}$ with Hermitian Hamiltonian $K_{\mathrm{eff}}(t_0) + \Delta(\rho_0)$ and jump operator $M(\rho_0) = \sqrt{\gamma(\rho_0)}\,L_+$ admitting a standard Stinespring dilation. Consequently, the resulting semiflow preserves trace normalization and density-matrix positivity ($\rho(t) \geq 0$) along the entire trajectory without generating negative probabilities.

\subsubsection{Resolution of Initial-Correlation Constraints via Pechukas's Theorem}

A secondary conceptual objection to state-dependent open-system dynamics stems from Pechukas's theorem~\cite{Pechukas1994}, which proves that a reduced dynamical map $\rho(t) = \mathcal{E}_t(\rho(0))$ admits a linear, state-independent assignment representation if and only if the global system--environment state begins in a strictly uncorrelated product state, $\rho_{\mathrm{total}}(0) = \rho(0) \otimes \rho_E(0)$.

In the non-equilibrium setting of an autonomous limit cycle, this equilibrium product-state assumption is explicitly violated. When an open system is coupled to a macroscopically synchronized phase background $\rho_E(t)$ supported on a $U(1)$ limit cycle, the system and environment maintain persistent phase correlations throughout their dynamical history. Along the Keldysh contour, these historical correlations are encoded directly in the Normalization Shield:
\begin{equation}
\mathcal{N}_{\mathrm{shield}} = \exp\left[-2\lambda \int_0^T dt \sum_{j=1}^2 \left(\mathcal{R}_j^+(t)\cos\big(\Phi_j^+(t) - \Psi_E(t)\big) - \mathcal{R}_j^-(t)\cos\big(\Phi_j^-(t) - \Psi_E(t)\big)\right)\right].
\end{equation}
Because the system's transition currents continuously entrain and recoil against the environmental phase field $\Psi_E(t)$, the global state evolves on a phase-entrained submanifold rather than a factorized tensor-product space. Projecting the global unitary evolution onto this correlated submanifold necessarily yields a state-dependent reduced generator---generating both the coherent recoil $\Delta(\rho)$ and the dissipative saturation $\gamma_q(\rho)$ as physical signatures of phase entrainment rather than mathematical artifacts.

\subsection{Bath-Factorized Closure and Quantum Causality}

Finally, we address concerns that nonlinear master equations can permit superluminal signaling. In ad-hoc nonlinear modifications of unitary quantum mechanics (such as Weinberg-type nonlinearities), state-dependent terms are postulated as fundamental modifications of single-particle evolution, allowing local state measurements to transmit instantaneous signals across spacelike-separated entangled subsystems.

In our framework, by contrast, both $\Delta(\rho)$ and $-\Phi_1|\mathrm{Tr}(\rho L_-)|^2\mathcal{D}_{L_+}[\rho]$ arise strictly as an effective bath-factorized closure in the thermodynamic many-body limit ($N_{\mathrm{bath}} \to \infty$) of the surrounding non-equilibrium environment, analogous to the time-dependent Kohn--Sham equations in density functional theory or the Gross--Pitaevskii equation for Bose--Einstein condensates. Here, $\Delta(\rho)$ and $\gamma(\rho)$ model the local physical back-reaction of the system onto a macroscopic environmental order parameter $Z(t) = \mathrm{Tr}_E(\rho_E(t)e^{ix})$ rather than a fundamental nonlinear law of closed-system quantum mechanics.

Furthermore, when extending the system to include an isolated ancilla $\mathcal{H}_A$, the environmental phase bridge couples locally to the system transition operators $L_\pm^{(j)} \otimes \mathbb{I}_A$. Evaluating the traced coherence amplitude on the extended state $\rho_{SA}$ gives $\mathrm{Tr}_{SA}[\rho_{SA}(L_-^{(j)} \otimes \mathbb{I}_A)] = \mathrm{Tr}_S[\rho_S L_-^{(j)}]$, which depends solely on the local reduced density matrix $\rho_S = \mathrm{Tr}_A(\rho_{SA})$. Consequently, local operations performed on a distant ancilla $\mathcal{H}_A$ cannot alter either $\Delta(\rho_S)$ or $\gamma_j(\rho_S)$, strictly preserving no-signaling and relativistic causality.

\section{Explicit examples and Semiclassical Correspondence}
\label{sec:correspondence}

We now demonstrate both branches of the dimensional dichotomy in Theorem~\ref{thm:structure} via explicit, closed-form realizations. First, we solve the single-qubit and bipartite two-qubit models using the leading-order Keldysh generator of Corollary~\ref{cor:keldysh_truncation}, showing how $\Delta(\rho)$ sustains hyperbolic limit cycles, mutual Kuramoto phase locking, and Arnold-tongue entrainment in finite dimensions with global positivity. We then project the same generator onto bosonic Fock space to establish its exact correspondence with phenomenological synchronization models~\cite{Lee2013,Chia2020}.

\subsection{Finite-Dimensional Realizations: Single-Qubit and Bipartite Entrainment}

In finite-dimensional spin systems, population depletion naturally saturates the radial amplitude, so we set $\Phi_1 = 0$ and $\chi = 0$ in Corollary~\ref{cor:keldysh_truncation} (with $\hbar = 1$), guaranteeing unconditional global positivity on the entire state space without a restricted basin.

\subsubsection{Single-Qubit Limit Cycle and Arnold Tongue ($\mathcal{H} = \mathbb{C}^2$)}

With $L_\pm = \sigma_\pm$, $H_0 = \frac{\omega}{2}\sigma_z$, $\Delta(\rho) = g\langle\sigma_-\rangle\sigma_+ + g^*\langle\sigma_+\rangle\sigma_-$ ($g = \alpha - i\beta$, $\beta > 0$), and $H_{\mathrm{ent}}(t) = \lambda\big(Z_0(t)\sigma_- + Z_0^*(t)\sigma_+\big)$ where $Z_0(t) = R_E e^{i\Omega_0 t}$, Eq.~(\ref{eq:keldysh_corollary}) becomes
\begin{equation}
\frac{d\rho}{dt} = -i\big[H_0 + \Delta(\rho) + H_{\mathrm{ent}}(t),\, \rho\big] + \Phi_0 \mathcal{D}_{\sigma_+}[\rho] + \Gamma_2 \mathcal{D}_{\sigma_-}[\rho].
\label{eq:single_qubit}
\end{equation}
Defining $w(t) := \langle\sigma_-\rangle = R(t)e^{-i\phi(t)}$, $z(t) := \langle\sigma_z\rangle$, $\Gamma_{\mathrm{tot}} := \Phi_0 + \Gamma_2$, and $\Delta\Gamma := \Phi_0 - \Gamma_2$, the exact Bloch equations close without approximation:
\begin{align}
\dot{w} &= \left(-i\omega + (i\alpha + \beta)z - \frac{\Gamma_{\mathrm{tot}}}{2}\right)w + i\lambda Z_0^*(t) z, \\
\dot{z} &= \Delta\Gamma - \Gamma_{\mathrm{tot}} z - 4\beta R^2 - 4\lambda\,\mathrm{Im}\big[Z_0(t) w\big].
\end{align}

In the uncoupled limit ($Z_0 = 0$), writing $w = R e^{-i\phi}$ decouples the radial-inversion dynamics from the phase:
\begin{equation}
\dot{R} = \left(\beta z - \frac{\Gamma_{\mathrm{tot}}}{2}\right)R, \qquad \dot{z} = \Delta\Gamma - \Gamma_{\mathrm{tot}} z - 4\beta R^2, \qquad \dot{\phi} = \omega - \alpha z.
\end{equation}
The trivial fixed point $(R = 0,\, z_0 = \Delta\Gamma/\Gamma_{\mathrm{tot}})$ undergoes a supercritical Hopf bifurcation when $\beta z_0 > \Gamma_{\mathrm{tot}}/2$:
\begin{equation}
2\beta(\Phi_0 - \Gamma_2) > (\Phi_0 + \Gamma_2)^2.
\label{eq:qubit_threshold}
\end{equation}
Notice that Eq.~(\ref{eq:qubit_threshold}) strictly requires population inversion ($\Phi_0 > \Gamma_2$); if $\pi_1(\mathcal{M}) = 0 \implies \Phi_0 = 0$, the threshold cannot be met and the qubit collapses to $z_0 = -1$. Above threshold, setting $\dot{R} = \dot{z} = 0$ yields a unique periodic orbit with frequency $\omega_* = \omega - \alpha z_*$:
\begin{equation}
z_* = \frac{\Phi_0 + \Gamma_2}{2\beta}, \qquad R_* = \sqrt{\frac{1}{4\beta}\left((\Phi_0 - \Gamma_2) - \frac{(\Phi_0 + \Gamma_2)^2}{2\beta}\right)} > 0.
\end{equation}
Because $\Delta\Gamma \leq \Gamma_{\mathrm{tot}}$ and $0 < z_* < 1$, the squared Bloch radius satisfies
\begin{equation}
4R_*^2 + z_*^2 = \frac{\Delta\Gamma}{\beta} - \frac{\Gamma_{\mathrm{tot}}^2}{4\beta^2} \leq 1 - (1 - z_*)^2 < 1,
\end{equation}
proving that the limit cycle lies strictly inside the physical Bloch ball for all parameters above threshold. Linearizing $(\dot{R}, \dot{z})$ around $(R_*, z_*)$ yields the Jacobian
\begin{equation}
J(R_*, z_*) = \begin{pmatrix} 0 & \beta R_* \\ -8\beta R_* & -\Gamma_{\mathrm{tot}} \end{pmatrix},
\end{equation}
with $\mathrm{Tr}(J) = -\Gamma_{\mathrm{tot}} < 0$ and $\det(J) = 8\beta^2 R_*^2 > 0$, confirming that both radial eigenvalues have strictly negative real parts and the limit cycle is hyperbolic and globally attracting for all $R(0) > 0$.

Switching on weak entrainment ($Z_0(t) = R_E e^{i\Omega_0 t}$) and projecting the relative phase $\psi(t) := \phi(t) - \Omega_0 t$ onto the limit-cycle shell $(R_*, z_*)$ yields the exact Adler equation,
\begin{equation}
\dot{\psi} = \Delta\omega - \frac{\lambda R_E z_*}{R_*}\cos\psi, \qquad \Delta\omega := (\omega - \alpha z_*) - \Omega_0,
\end{equation}
which produces stable phase locking ($\sin\psi_* < 0$) inside the exact Arnold tongue $|\Delta\omega| \leq \lambda R_E z_* / R_*$. Adding any linear Lindblad channel (such as pure dephasing $\gamma_\phi\mathcal{D}_{\sigma_z}[\rho]$) preserves global positivity unconditionally and simply shifts the threshold to $2\beta(\Phi_0 - \Gamma_2) > (\Phi_0 + \Gamma_2)(\Phi_0 + \Gamma_2 + 4\gamma_\phi)$.

\begin{figure}[h]
    \centering
    \includegraphics[width=0.92\textwidth]{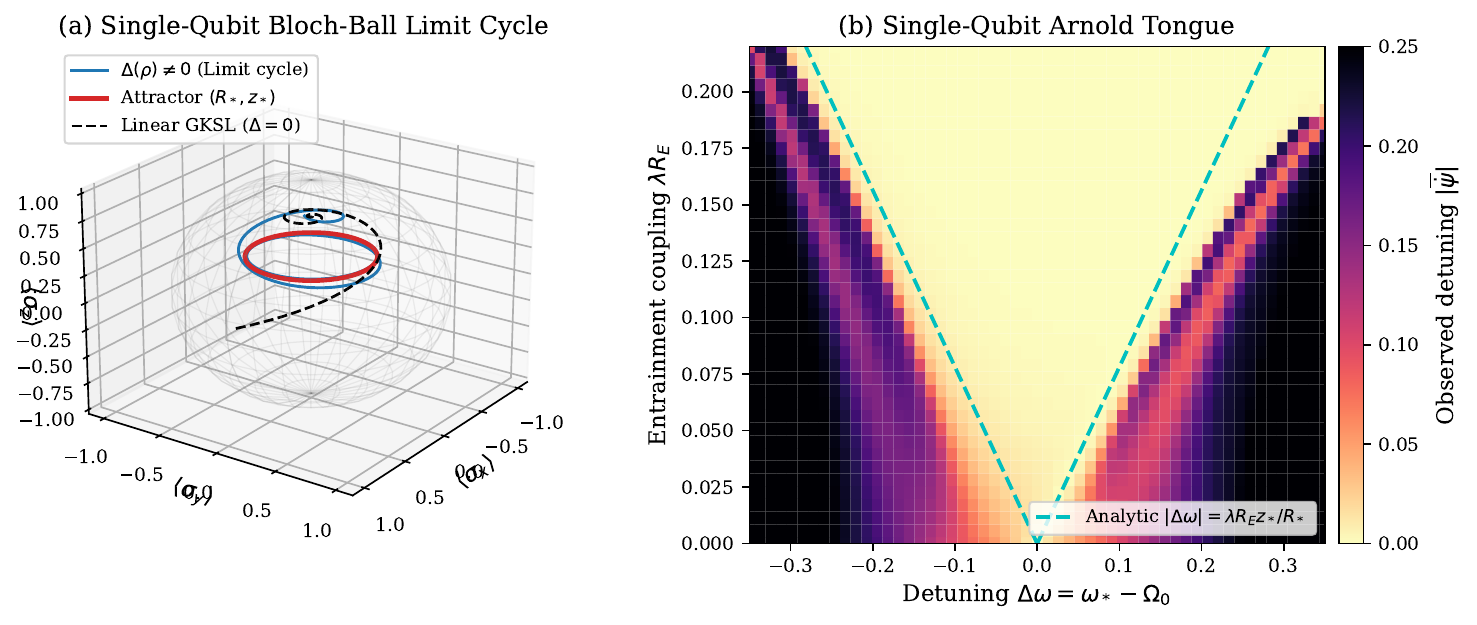}
    \caption{\textbf{Single-qubit autonomous limit cycle and entrainment.}
    (a) Exact Bloch-ball trajectories at $Z_0 = 0$ ($\Phi_0 = 1.0$, $\Gamma_2 = 0.2$, $\alpha = 0.5$, $\beta = 2.0$): with coherent recoil $\Delta(\rho) \neq 0$, trajectories spiral onto the hyperbolic limit-cycle shell $(R_*, z_*)$ strictly inside the Bloch ball, whereas the linear GKSL limit ($\Delta = 0$) collapses unconditionally to the $z$-axis fixed point.
    (b) Numerical Arnold tongue showing the time-averaged phase drift $|\overline{\dot{\psi}}|$ in the $(\Delta\omega, \lambda R_E)$ plane, matching the exact analytical Adler boundary $|\Delta\omega| = \lambda R_E z_*/R_*$ (dashed cyan).}
    \label{fig:single_qubit}
\end{figure}

\subsubsection{Bipartite Two-Qubit Mutual Phase Locking ($\mathcal{H} = \mathbb{C}^2 \otimes \mathbb{C}^2$)}

Coupling two qubits ($j \in \{1, 2\}$) with $H_{\mathrm{bi}} = \frac{\omega_1}{2}\sigma_z^{(1)} + \frac{\omega_2}{2}\sigma_z^{(2)}$ to the shared $U(1)$ bath via Eq.~(\ref{eq:coherent_recoil}) generates the collective coherent recoil $\Delta(\rho) = \sum_{j,k=1}^2\big(g\langle\sigma_-^{(k)}\rangle\sigma_+^{(j)} + \mathrm{h.c.}\big)$ and the bipartite master equation
\begin{equation}
\frac{d\rho}{dt} = -i\big[H_{\mathrm{bi}} + \Delta(\rho) + H_{\mathrm{ent}}(t),\, \rho\big] + \sum_{j=1}^2\Big(\Phi_0\mathcal{D}_{\sigma_+^{(j)}}[\rho] + \Gamma_2\mathcal{D}_{\sigma_-^{(j)}}[\rho]\Big).
\end{equation}
Because $\langle\sigma_-^{(k)}\rangle$ enters $\Delta(\rho)$ as a scalar expectation value, the coupled Bloch equations for $w_j := \langle\sigma_-^{(j)}\rangle = R_j e^{-i\phi_j}$ and $z_j := \langle\sigma_z^{(j)}\rangle$ close exactly:
\begin{align}
\dot{w}_j &= \left(-i\omega_j - \frac{\Gamma_{\mathrm{tot}}}{2}\right)w_j + (i\alpha + \beta)z_j(w_1 + w_2) + i\lambda Z_0^*(t) z_j, \\
\dot{z}_j &= \Delta\Gamma - \Gamma_{\mathrm{tot}} z_j - 4\beta R_j^2 + 4\,\mathrm{Im}\big[g w_j^* w_k\big] - 4\lambda\,\mathrm{Im}\big[Z_0(t) w_j\big] \qquad (k \neq j).
\end{align}
At $Z_0 = 0$, identical qubits ($\omega_1 = \omega_2 = \omega$) form a symmetric collective limit cycle ($R_1 = R_2 = R$, $\phi_1 = \phi_2 = \phi$, $z_1 = z_2 = z$) with the threshold halved by collective feedback, $4\beta(\Phi_0 - \Gamma_2) > (\Phi_0 + \Gamma_2)^2$, and physical orbit $z_* = (\Phi_0 + \Gamma_2)/(4\beta)$, $R_* = \sqrt{\frac{1}{8\beta}[(\Phi_0 - \Gamma_2) - (\Phi_0 + \Gamma_2)^2/(4\beta)]}$ satisfying $4R_*^2 + z_*^2 < 1$.

For detuned qubits ($\delta\omega := \omega_1 - \omega_2 \neq 0$), extracting the phase equations from $\dot{w}_j$ on the symmetric shell ($R_1 \approx R_2 \approx R_*$, $z_1 \approx z_2 \approx z_*$) cancels the reactive $\alpha\cos\Delta\phi$ terms identically and yields the exact Kuramoto mutual phase-locking equation for $\Delta\phi := \phi_1 - \phi_2$:
\begin{equation}
\frac{d(\Delta\phi)}{dt} = (\omega_1 - \omega_2) - 2\beta z_*\sin(\Delta\phi) = (\omega_1 - \omega_2) - \frac{\Phi_0 + \Gamma_2}{2}\sin(\Delta\phi).
\end{equation}
For any detuning $|\omega_1 - \omega_2| \leq (\Phi_0 + \Gamma_2)/2$, the two qubits lock mutually at $\Delta\phi_* = \arcsin[2(\omega_1 - \omega_2)/(\Phi_0 + \Gamma_2)]$ with negative phase-difference eigenvalue $-\frac{1}{2}(\Phi_0 + \Gamma_2)\cos\Delta\phi_* < 0$. Under weak $Z_0(t) \neq 0$, the locked mean phase $\bar{\psi} := \frac{1}{2}(\phi_1 + \phi_2) - \Omega_0 t$ obeys the collective Adler equation $\dot{\bar{\psi}} = \bar{\Delta\omega} - (\lambda R_E z_*/R_*)\cos\bar{\psi}$, entraining the bipartite pair to $\Omega_0$.

\begin{figure}[h]
    \centering
    \includegraphics[width=\textwidth]{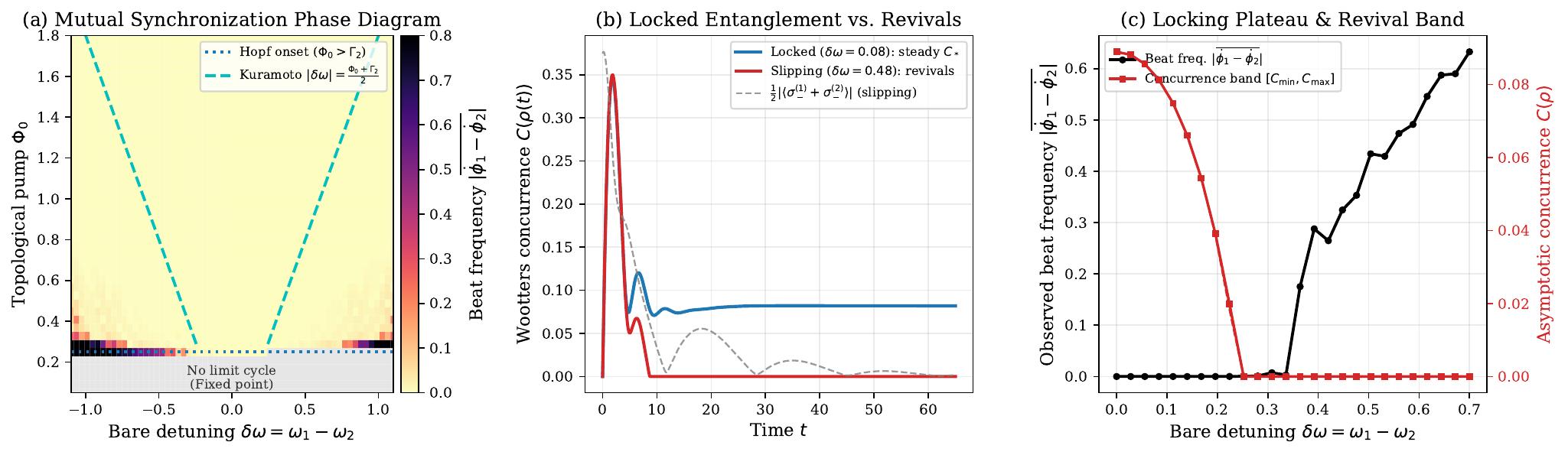}
    \caption{\textbf{Bipartite mutual synchronization and periodic concurrence revivals.}
    (a) Mutual synchronization phase diagram in the $(\delta\omega, \Phi_0)$ plane ($\Gamma_2 = 0.2$, $\alpha = 0.3$, $\beta = 1.0$, $J_{zz} = 0$), showing the time-averaged beat frequency $|\overline{\dot{\phi}_1 - \dot{\phi}_2}|$. Gray marks the sub-threshold fixed-point region below the collective Hopf onset (dotted blue, confirming $\Phi_0 > \Gamma_2$ from Theorem~\ref{thm:inverse_topology}); above threshold, the mutual locking tongue matches the exact Kuramoto boundary $|\delta\omega| = (\Phi_0 + \Gamma_2)/2$ (dashed cyan).
    (b) Exact $4\times 4$ density-matrix evolution of Wootters concurrence $C(\rho(t))$ from a separable product state with $U(1)$-covariant coupling $J_{zz} = 0.35$ ($\Phi_0 = 0.40$, $\Gamma_2 = 0.01$, $\beta = 0.16$): inside the locking tongue ($\delta\omega = 0.08$, blue), synchronization stabilizes steady-state entanglement, whereas outside the tongue ($\delta\omega = 0.48$, red), phase slips modulate the collective coherence $|\langle\sigma_-^{(1)} + \sigma_-^{(2)}\rangle|$ (dashed gray) and drive sustained periodic concurrence revivals.
    (c) Frequency-pulling plateau (black) overlaid with the asymptotic concurrence band $[C_{\min}, C_{\max}]$ (red shading) across the Kuramoto unlocking transition.}
    \label{fig:bipartite_sync}
\end{figure}

Because the bare recoil $\Delta(\rho)$ and local dissipators act additively across the two qubits, an initially separable state remains unentangled in the absence of direct inter-qubit interactions. Adding a $U(1)$-covariant Ising coupling $J_{zz}\sigma_z^{(1)}\sigma_z^{(2)}$ to $ H_{\mathrm{bi}}$ which still remains strictly within the class $\mathfrak{G}_{\mathrm{osc}}$ of Theorem~\ref{thm:structure} while coupling the Kuramoto synchronization transition directly to bipartite entanglement. As shown in Fig.~\ref{fig:bipartite_sync}, mutual phase locking inside the Kuramoto tongue sustains constructive collective recoil $|w_1 + w_2| > 0$, stabilizing a stationary entangled plateau ($C(\rho) \to C_* > 0$) from an unentangled product state. Outside the locking tongue ($|\delta\omega| > \delta\omega_c$), periodic phase slips alternately quench ($w_1 + w_2 \approx 0$ at $\Delta\phi \approx \pi$) and reignite the collective coherence, driving sustained periodic concurrence revivals synchronized to the beat frequency.

\subsection{Bosonic Fock-Space Correspondence with Phenomenological Models}

In the infinite-dimensional bosonic regime ($\dim\mathcal{H} = \aleph_0$, $\Delta(\rho) = 0$), Corollary~\ref{cor:keldysh_truncation} establishes a direct correspondence with phenomenological models of quantum synchronization~\cite{Lee2013,Chia2020}, which construct limit cycles top-down by postulating a linear single-photon pump balanced against nonlinear two-photon loss. Specifically, Lee \textit{et al.}~\cite{Lee2013} begin with the weakly nonlinear classical van der Pol oscillator, $\ddot{x} - \epsilon(1 - x^2)\dot{x} + \omega_0^2 x = 0$ ($\epsilon \ll 1$), which under the rotating-wave ansatz $x(t) = \alpha(t)e^{-i\omega_0 t} + \mathrm{c.c.}$ reduces to the Stuart--Landau amplitude equation,
\begin{equation}
\frac{d\alpha}{dt} = -i\omega_0 \alpha + \frac{\epsilon}{2}\big(1 - |\alpha|^2\big)\alpha.
\end{equation}
To reproduce this drift within a strictly linear GKSL master equation, they postulate
\begin{equation}
\frac{d\rho_{\mathrm{phen}}}{dt} = -i[\omega_0 a^\dagger a, \rho_{\mathrm{phen}}] + \kappa_1 \mathcal{D}_{a^\dagger}[\rho_{\mathrm{phen}}] + \kappa_2 \mathcal{D}_{a^2}[\rho_{\mathrm{phen}}].
\end{equation}

Corollary~\ref{cor:keldysh_truncation} reverses this logic. Projecting Eq.~(\ref{eq:keldysh_corollary}) onto a single active bosonic charge sector ($q = 1$, $L_-^{(1)} \mapsto a$, $L_+^{(1)} \mapsto a^\dagger$, $\Delta(\rho) = 0$, $\Gamma_2^{(1)} = 0$, $\mathcal{L}_{\mathrm{lin}} = 0$) yields the exact first-moment Stuart--Landau evolution for $z_1 = \langle a \rangle$:
\begin{enumerate}
\item \textbf{First-Moment Closure from Eq.~(\ref{eq:keldysh_corollary}):}
\begin{equation}
\frac{d\langle a \rangle}{dt} = \mathrm{Tr}\!\left(a\,\mathcal{L}(\rho)\right) = -i\Omega_0 \langle a \rangle + \frac{1}{2}\Big(\Phi_0^{(1)} - \Phi_1^{(1)}|\langle a \rangle|^2\Big)\langle a \rangle,
\end{equation}
which closes \emph{without approximation} across all excitation regimes.
\item \textbf{Unclosed Moment Hierarchy in $\mathcal{L}_{\mathrm{phen}}$:}
\begin{equation}
\frac{d\langle a \rangle}{dt} = \mathrm{Tr}\!\left(a\,\mathcal{L}_{\mathrm{phen}}[\rho]\right) = -i\omega_0 \langle a \rangle + \frac{\kappa_1}{2}\langle a \rangle - \kappa_2 \langle a^\dagger a^2 \rangle.
\end{equation}
Matching this to the Stuart--Landau drift requires a semiclassical cumulant truncation around $a = \langle a \rangle + \delta a$, $\langle a^\dagger a^2 \rangle \approx |\langle a \rangle|^2 \langle a \rangle$, dropping $O(1/|\alpha|^2)$ fluctuation cumulants and fixing $\kappa_1 \equiv \Phi_0^{(1)}$, $\kappa_2 \equiv \Phi_1^{(1)}/2$.
\end{enumerate}

\begin{remark}[Extension to Non-Circular Relaxation Oscillations]
The same reverse-engineering mechanism underlies subsequent extensions to strongly nonlinear relaxation oscillations~\cite{Chia2020}, where non-circular phase-space trajectories (such as the full Li\'enard system without rotating-wave averaging) are matched at the Wigner-function level by synthesizing quadrature-dependent jump operators such as $L_2 \propto \hat{x}a$. More generally, pushing the radial flow forward under a smooth coordinate map $(x,p) = \Psi(z_1, z_1^*)$ and reconstructing it inside a strictly linear GKSL template requires a hierarchical cascade of polynomial jump operators $L_k(\hat{x}, \hat{p})$, recovering the cascade quantization formalism~\cite{Chia2025}.
\end{remark}

\subsection{Resolution of the Multi-Photon Scattering Puzzle}

In a physical laboratory, realizing a two-photon Lindblad dissipator $\mathcal{D}_{a^2}[\rho]$ requires engineering exotic two-quantum reservoir couplings that are forbidden at leading order by the $\Delta n = \pm 1$ winding selection rule of Theorem~\ref{thm:inverse_topology}. Our reduction demonstrates that \emph{two-photon dissipation is not a fundamental microscopic process, but a linearized GKSL surrogate for single-quantum gain saturation}:
\begin{enumerate}
\item \textbf{Microscopic Picture (Eq.~(\ref{eq:keldysh_corollary})):} The bosonic limit cycle is stabilized entirely within the single-quantum channel ($\Delta n = \pm 1$). As coherence builds along the $U(1)$ phase bridge, the Normalization Shield recoil ($-\Phi_1^{(1)}|z_1|^2$) smoothly quenches the upward pumping rate $\gamma_1^+(\rho)$ until it balances linear damping at $|z_{1,*}|^2 = (\Phi_0^{(1)} - \Gamma_2^{(1)})/\Phi_1^{(1)}$.
\item \textbf{Phenomenological Approximation ($\mathcal{L}_{\mathrm{phen}}$):} Insisting on a state-independent linear Lindblad equation forbids gain quenching; instead, $\kappa_1\mathcal{D}_{a^\dagger}[\rho]$ is kept fixed at its unsaturated value and $\kappa_2\mathcal{D}_{a^2}[\rho]$ is introduced as a linear surrogate whose semiclassical drift reproduces $-\frac{1}{2}\Phi_1^{(1)}|z_1|^2 z_1$.
\end{enumerate}
Consequently, while $\mathcal{L}_{\mathrm{phen}}$ captures the macroscopic limit-cycle radius in the large-amplitude semiclassical regime ($|\alpha|^2 \gg 1$), it relaxes to a phase-diffused stationary state ($\langle a \rangle = 0$) and introduces spurious two-photon noise in the deep quantum regime ($|\alpha| \lesssim 1$), whereas Eq.~(\ref{eq:keldysh_corollary}) preserves phase coherence and exact first-moment closure.

\section{Conclusion}
\label{sec:conclusions}

This work has established a unified geometric and operator-algebraic framework for open quantum systems that sustain phase-coherent, entrainable limit-cycle attractors across arbitrary separable Hilbert spaces ($\dim\mathcal{H} \leq \aleph_0$). Replacing the classical 1-form action with a Poincar\'e--Cartan 2-form integrated over an $\epsilon$-filament surface in extended phase space resolves the inability of conservative path integrals to describe autonomous phase-space contraction. The resulting non-self-adjoint Gelfand fluctuation operator generates a real exponential amplitude-focusing envelope that concentrates the quantum wave packet onto a stable periodic orbit.

Evaluating this 2-form path integral for a quantum system coupled to a self-oscillating $U(1)$ phase environment collapses continuous spatial integration into discrete integer winding jumps. The asymmetric boundary phase terms compile into the trace-preserving Normalization Shield ($\mathcal{N}_{\mathrm{shield}}\vert{}_{\mathrm{diag}} = 1$), while the system's reciprocal back-action on the bath order parameter simultaneously generates the $O(\lambda^2)$ coherent recoil $\Delta(\rho)$ and the $O(\lambda^4)$ dissipative gain-saturation rate $-\Phi_1\vert{}\mathrm{Tr}(\rho L_-)\vert{}^2\mathcal{D}_{L_+}[\rho]$.

Two central theorems anchor this framework:
\begin{enumerate}
\item \textbf{Inverse Coupling Theorem (Theorem~\ref{thm:inverse_topology}):} Sustaining an autonomous, entrainable quantum limit cycle requires an interaction manifold $\mathcal{M}$ with a non-trivial first homotopy group $\pi_1(\mathcal{M}) = \mathbb{Z}$ and active winding pumping above linear damping ($\Phi_0 > \Gamma_2$), uniquely selecting the compact circle group $S^1 \cong U(1)$. Simply connected manifolds ($\pi_1(\mathcal{M}) = 0$) extinguish the winding pump ($\Phi_0 = 0$) and force relaxation to stationary fixed points in both finite and infinite dimensions.
\item \textbf{Structure Theorem (Theorem~\ref{thm:structure}):} Under  $U(1)$ covariance, a symmetric bidirectional phase interface within the frozen-state GKSL cone, and an entrainable hyperbolic $S^1$ limit cycle, the reduced generator must take the universal operator form of Eq.~(\ref{eq:universal_structure}). Density-matrix positivity enforces a sharp dimensional dichotomy: purely dissipative gain saturation is strictly forbidden in finite dimensions ($\dim\mathcal{H} < \infty$) by trace-norm contraction, making coherent recoil $\Delta(\rho)$ mandatory for qubits and qudits, whereas unbounded ladder algebras ($\dim\mathcal{H} = \aleph_0$) permit purely dissipative saturation. Expanding the feedback functionals in the  coupling yields the leading-order polynomial master equation (Corollary~\ref{cor:keldysh_truncation}) as well as a systematic hierarchy for subcritical regimes and Pad\'e-resummed laser saturation.
\end{enumerate}

Explicit single-qubit and bipartite two-qubit realizations verify that $\Delta(\rho)$ sustains globally positive hyperbolic limit cycles, exact Kuramoto mutual phase locking, and Arnold-tongue entrainment in finite dimensions---resolving the no-go obstruction for linear GKSL qubits~\cite{Roulet2018}---while projecting the same generator onto bosonic Fock space recovers phenomenological quantum van der Pol models as linearized surrogates for single-quantum gain saturation~\cite{Lee2013,Chia2020}.

\paragraph*{Outlook}
The geometric path-integral framework developed here opens several clear directions for future investigation:
\begin{enumerate}
\item \textbf{Many-Body Network Synchronization:} Extending the bipartite master equation to an arbitrary $N$-body network of nonlinear quantum nodes enables the study of multi-node phase locking, chimera states, and topological boundary synchronization in non-equilibrium open quantum arrays.
\item \textbf{Quantum Time Crystals:} Exploring the connection between the discrete winding sectors of the 2-form path integral and spontaneous discrete time-translation symmetry breaking provides a first-principles foundation for classifying open, driven-dissipative quantum time crystals.
\item \textbf{Higher-Order $p:q$ Entrainment in Qudits:} Applying the higher-order cumulant hierarchy of Eq.~(\ref{eq:keldysh_hierarchy}) to multi-level qudit systems ($d \geq 3$) will enable the analysis of subcritical Hopf transitions, bistable limit cycles, and $p:q$ harmonic Arnold tongues on complex projective manifolds.
\end{enumerate}

Beyond periodic attractors, the Spectral Mechanics framework of \cite{Siddhartha2026} naturally accommodates environments supporting richer non-equilibrium dynamics, including energy cascades, quasi-periodic tori, and chaotic attractors. Extending our classification to these regimes is left for future investigation.

\appendix
\section{Infinite-Dimensional Domain Regularity, Invariance, and Channel Addition}
\label{app:domain_regularity}

In this appendix, we prove that our limit-cycle master equation remains mathematically well-posed in countably infinite dimensions ($\dim\mathcal{H} = \aleph_0$), such as bosonic oscillators where the ladder operators $L_\pm^{(j)}$ are unbounded. We restrict to physical states with finite average energy and establish four lemmas: (1) finite-energy control, (2) trapping inside the positive-rate basin $\mathcal{B}_{1,2}$, (3) valid frozen-state CPTP evolution at every instant, and (4) safe composition with linear Lindblad channels.

For finite-dimensional systems ($\dim\mathcal{H} < \infty$) with $\Phi_1 = 0$, all operators are bounded and $\gamma_q^\pm \geq 0$ globally; existence, uniqueness, trace preservation, positivity, and composition with arbitrary linear Lindblad channels hold unconditionally on the entire state space by standard matrix ODE theory.

\subsection{Setup: Finite-Energy States and Core Subspaces}

To handle unbounded transition operators $L_\pm^{(j)}$ (such as bosonic creation and annihilation operators $a_j^\dagger, a_j$), we measure the excitation level of each subsystem using a positive number operator $N_j \geq \mathbb{I}$ with integer spectrum $\{1, 2, 3, \dots\}$ and define the combined weight $W = N_1 \otimes N_2$. The space of \textbf{finite-energy states} $\mathfrak{T}_{1,2}$ consists of all trace-class operators $\rho$ with finite weighted trace norm,
\begin{equation}
\|\rho\|_{1,2} := \mathrm{Tr}\!\left(W |\rho| W\right) < \infty,
\end{equation}
which satisfies $\|\rho\|_1 \leq \|\rho\|_{1,2}$. The \textbf{finite-level core} $\mathcal{D}_0 \subset \mathfrak{T}_{1,2}$ is the dense subspace of states supported on finitely many energy levels. Throughout this appendix, we assume two standard properties satisfied by bosonic ladder algebras:
\begin{enumerate}
\item[(A1)] \textit{Single-Step Shifts:} Applying $L_\pm^{(j)}$ shifts the excitation level by at most one step, and the weighted operators $N_j^{-1/2} L_\pm^{(j)} N_j^{-1/2}$ are bounded by a finite constant $C_j > 0$.
\item[(A2)] \textit{Ladder Commutator:} On $\mathcal{D}_0$, the transition operators satisfy $[L_-^{(j)}, L_+^{(j)}] = \tilde{\Delta}_j$, where $\tilde{\Delta}_j > 0$ is a positive operator ($\tilde{\Delta}_j = \mathbb{I}$ for bosons) bounded by a multiple of $N_j$.
\end{enumerate}

\subsection{Infinite-Dimensional Lemmas}

\begin{lemma}[Well-Behaved Coherence and Finite-Energy Evolution]
\label{lem:energy_bound}
On $\mathfrak{T}_{1,2}$, the coherence amplitude $R_j(\rho) = |\mathrm{Tr}(\rho L_-^{(j)})|$ is continuous and bounded by $C_j \|\rho\|_{1,2}$. Furthermore, any trajectory starting with finite energy retains finite energy for all $t \geq 0$ and defines a unique solution $\rho(t) \in \mathfrak{T}_{1,2}$.
\end{lemma}

\begin{proof}
Using the cyclic property of the trace and Assumption~(A1),
\begin{equation}
R_j(\rho) = \big|\mathrm{Tr}\big(W \rho W\, W^{-1} L_-^{(j)} W^{-1}\big)\big| \leq \big\|W^{-1} L_-^{(j)} W^{-1}\big\|_{\mathrm{op}}\|\rho\|_{1,2} \leq C_j \|\rho\|_{1,2}.
\end{equation}
Because $R_j(\rho)$ is continuous in $\|\cdot\|_{1,2}$, the positive-rate basin $\mathcal{B}_{1,2} = \{\rho \in \mathfrak{T}_{1,2} : \mathrm{Tr}(\rho) = 1,\; R_j^2(\rho) \leq \Phi_0/\Phi_1\}$ is closed and convex. Evaluating $\frac{d}{dt}\mathrm{Tr}(\rho W^2)$ on $\mathcal{D}_0$ via Assumption~(A1) shows that each Lindblad jump and coherent commutator shifts the energy weight at most linearly:
\begin{equation}
\frac{d}{dt}\mathrm{Tr}\big(\rho(t) W^2\big) \leq c_1 \mathrm{Tr}\big(\rho(t) W^2\big) + c_2
\end{equation}
for fixed constants $c_1, c_2 > 0$. By Gronwall's inequality, $\|\rho(t)\|_{1,2} \leq (\|\rho(0)\|_{1,2} + c_2/c_1)e^{c_1 t} < \infty$ for all finite $t \geq 0$, guaranteeing a unique trajectory in $\mathfrak{T}_{1,2}$.
\end{proof}

\begin{lemma}[Trapping Inside the Positive-Rate Basin]
\label{lem:invariance}
In the truncated polynomial regime ($\Phi_1 > 0$), if $\rho(0) \in \mathcal{B}_{1,2}$ and the entrainment and cross-talk couplings satisfy $2\lambda R_E < \Gamma_2 \delta_0 \sqrt{\Phi_0/\Phi_1}$ and $2\chi\sum_{k \neq j} R_k < \Phi_0$, then $\rho(t) \in \mathcal{B}_{1,2}$ for all $t \geq 0$. Consequently, $\gamma_j(\rho(t)) \geq 0$ for all $t \geq 0$.
\end{lemma}

\begin{proof}
Consider a finite-level state $\rho \in \mathcal{D}_0$ on the boundary $\partial\mathcal{B}_{1,2}$, where $\gamma_j(\rho) = 0$ (reducing to $R_j^2(\rho) = \Phi_0/\Phi_1$ when $\chi = 0$). On $\partial\mathcal{B}_{1,2}$, the unbounded pumping channel $\gamma_j(\rho)\mathcal{D}_{L_+^{(j)}}[\rho]$ vanishes identically, leaving only the coherent terms and the contractive damping $\Gamma_2\mathcal{D}_{L_-^{(j)}}[\rho] + \mathcal{L}_{T,j}[\rho]$. Using $[L_-^{(j)}, L_+^{(j)}] = \tilde{\Delta}_j > 0$ (Assumption~(A2)), the bare Hamiltonian $H_0$ preserves $R_j^2$, while $H_{\mathrm{ent}}(t)$ contributes at most $+2\lambda R_E \sqrt{\Phi_0/\Phi_1}$ and linear damping pulls strictly inward:
\begin{equation}
\left.\frac{d}{dt}R_j^2(\rho)\right|_{\partial\mathcal{B}_{1,2}} \leq -\big(\Gamma_2 \delta_0 + \gamma_{\mathrm{th}}\big)\frac{\Phi_0}{\Phi_1} + 2\lambda R_E \sqrt{\frac{\Phi_0}{\Phi_1}} < 0,
\end{equation}
where $\delta_0 > 0$ is the spectral lower bound of $\tilde{\Delta}_j$ and $\gamma_{\mathrm{th}} \geq 0$ is the thermal damping rate. Since $\mathcal{D}_0 \cap \partial\mathcal{B}_{1,2}$ is dense in $\partial\mathcal{B}_{1,2}$ under $\|\cdot\|_{1,2}$, Nagumo's invariance theorem~\cite{Nagumo1942} guarantees that no trajectory starting in $\mathcal{B}_{1,2}$ can exit. (When $\gamma_j^+$ is globally positive, e.g., under Pad\'e resummation, no boundary $\partial\mathcal{B}_{1,2}$ exists and invariance holds on all of $\mathfrak{T}_{1,2}$.)
\end{proof}

\begin{lemma}[Valid Quantum Evolution at Every Instant]
\label{lem:tangent_cptp}
For any frozen state $\rho_0 \in \mathcal{B}_{1,2}$, the frozen tangent generator $\mathcal{L}_{\rho_0}$ defines a valid, completely positive and trace-preserving (CPTP) quantum evolution on $\mathcal{H}$.
\end{lemma}

\begin{proof}
By Lemma~\ref{lem:invariance}, any $\rho_0 \in \mathcal{B}_{1,2}$ satisfies $0 \leq \gamma_j(\rho_0) \leq \Phi_0$, and $K_{\mathrm{eff}}(t_0) + \Delta(\rho_0)$ is Hermitian. Freezing the coefficients at $\rho_0$ yields a linear time-independent GKSL generator with jump operators $M_j = \sqrt{\gamma_j(\rho_0)}\,L_+^{(j)}$ and $\sqrt{\Gamma_2}\,L_-^{(j)}$. Because $\sqrt{\gamma_j(\rho_0)} \leq \sqrt{\Phi_0}$, $M_j$ satisfies the same linear energy-growth bound of Lemma~\ref{lem:energy_bound} with constants independent of $\rho_0$. By Davies' theorem for unbounded Lindblad generators~\cite{Davies1977,Chebotarev1998}, probability is conserved ($\mathrm{Tr}[\mathcal{L}_{\rho_0}(\rho)] = 0$) and $e^{t\mathcal{L}_{\rho_0}}$ is a CPTP contraction admitting a standard Stinespring dilation.
\end{proof}

\begin{lemma}[Safe Composition with Linear Lindblad Channels]
\label{lem:trotter_addition}
Adding any linear Lindblad generator $\mathcal{L}_{\mathrm{lin}}$ satisfying the linear energy-growth bound of Lemma~\ref{lem:energy_bound} (and, in the truncated polynomial regime $\Phi_1 > 0$, not inducing an outward radial drift exceeding $\Gamma_2\delta_0\frac{\Phi_0}{\Phi_1}$ on $\partial\mathcal{B}_{1,2}$) to $\mathcal{L}_{\mathrm{top}}$ produces a unique, trace-preserving, positivity-preserving evolution given by the Lie--Trotter product formula:
\begin{equation}
\rho(t) = \lim_{n \to \infty}\left(e^{\frac{t}{n}\mathcal{L}_{\mathrm{top}}(\cdot)} \circ e^{\frac{t}{n}\mathcal{L}_{\mathrm{lin}}}\right)^n \rho(0).
\end{equation}
\end{lemma}

\begin{proof}
Taking alternating steps of size $\tau = t/n$, each linear CPTP step $e^{\tau\mathcal{L}_{\mathrm{lin}}}$ preserves positivity, trace, and $\mathfrak{T}_{1,2}$ while respecting the inward drift on $\partial\mathcal{B}_{1,2}$; each subsequent step $e^{\tau\mathcal{L}_{\mathrm{top}}(\cdot)}$ preserves positivity, trace, and $\mathcal{B}_{1,2}$ with uniformly bounded $\|\rho\|_{1,2}$ by Lemmas~\ref{lem:energy_bound}--\ref{lem:tangent_cptp}. Expanding on $\mathcal{D}_0$ (where all operators are bounded and $\gamma_q, \Delta$ are Lipschitz continuous in $\|\cdot\|_{1,2}$ by Lemma~\ref{lem:energy_bound}) yields
\begin{equation*}
e^{\tau\mathcal{L}_{\mathrm{top}}(\cdot)}\!\left(e^{\tau\mathcal{L}_{\mathrm{lin}}}\rho\right) = \rho + \tau\big(\mathcal{L}_{\mathrm{lin}}(\rho) + \mathcal{L}_{\mathrm{top}}(\rho)\big) + O(\tau^2).
\end{equation*}
By the nonlinear Trotter--Kato product formula on Banach spaces~\cite{Brezis1972}, the alternating sequence converges uniformly in the trace norm $\|\cdot\|_1$ to the unique, trace-preserving, positivity-preserving solution of the combined master equation.
\end{proof}


\end{document}